\documentclass[12pt]{article}

\pdfoutput=1

\usepackage{graphicx}

\usepackage{amsmath}
\usepackage{color}
\begin{document}

\hoffset = -0.3truecm
\voffset = -1.1truecm

\title{\bf
Particles of One-Half Topological Charge\footnote{To be submitted for publication}}

\author{
{\bf Rosy Teh\footnote{E-mail: rosyteh@usm.my},}
{\bf Ban-Loong Ng and Khai-Ming Wong}\\
{\normalsize School of Physics, Universiti Sains Malaysia}\\
{\normalsize 11800 USM Penang, Malaysia}}

\date{August 7, 2012}
\maketitle

\begin{abstract}
We would like to show the existence of finite energy SU(2) Yang-Mills-Higgs particles of one-half topological charge. The magnetic fields of these solutions at spatial infinity correspond to the magnetic field of a positive one-half magnetic monopole located at the origin and a semi-infinite Dirac string which carries a magnetic flux of $\frac{2\pi}{g}$ going into the center of the sphere at infinity. Hence the net magnetic charge of the configuration is zero. The solutions possess gauge potentials that are singular along one-half of the $z$-axis, elsewhere they are regular. There are two distinct configurations of these particles with different total energies and magnetic dipole moments. Their total energies are found to increase with the strength of the Higgs field self-coupling constant $\lambda$. 
\end{abstract}


\section{Introduction}
The SU(2) Yang-Mills-Higgs (YMH) field theory in $3+1$ dimensions, with the Higgs field in the adjoint representation possess magnetic monopole configurations \cite{kn:1}-\cite{kn:4}. These monopole solution belongs to the category of solutions which are invariant under a U(1) subgroup of the local SU(2) gauge group. The  't Hooft-Polyakov monopole solution with non-zero Higgs mass and self-interaction belongs to such category and it is the first finite energy monopole solution. This numerical monopole solution of unit magnetic charge is spherically symmetric \cite{kn:1}. However, finite energy monopole configurations with magnetic charges greater than unity \cite{kn:4} cannot possess spherical symmetry \cite{kn:5}. Exact monopole solutions exist only in the Bogomol'nyi-Prasad-Sommerfield (BPS) limit \cite{kn:3}-\cite{kn:4}. Outside this limit and when the Higgs field potential is non-vanishing, only numerical solutions are known. Other exact solutions found are the A-M-A (antimonopole-monopole-antimonopole) and vortex ring solutions and various other mirror symmetric monopole configurations \cite{kn:6}.
Numerical BPS monopole solutions with no rotational symmetry have also been discussed \cite{kn:7}. Other numerical finite energy monopole solutions include the MAP (monopole-antimonopole-pair), MAC (monopole-antimonopole-chain), and vortex ring solutions \cite{kn:8}. 

However, most of these monopole solutions reported so far are of integer topological monopole charges. 
Hence it is our purpose in this paper to discuss the existence of finite energy one-half monopole magnetic charge solutions. These one-half monopole solutions are numerical solutions that are solved by first obtaining the exact one-half monopole solutions at large distances and then fixing the boundary conditions at small distances and along the $z$-axis. 

We analyse four types of one-half monopole solutions which we label as Type $A1$, $A2$, $B1$, and $B2$. Our analysis shows that the Type $B$ ($B1$ and $B2$) solutions are exactly $180^o$ rotation of the $z$-axis about the origin, $r=0$, of the Type $A$ ($A1$ and $A2$) solutions (respectively). However for a particular value of Higgs self-coupling constant, $\lambda$, the Type 1 ($A1$ and $B1$) and Type 2 ($A2$ and $B2$) solutions possess different total energies and magnetic dipole moments. Comparisons of the gauge invariant form of the gauge field potentials and Higgs fields are also made between the four one-half monopole solutions to confirm our conclusion that the Type 1 and Type 2 solutions are different one-half monopoles. 

The gauge potentials of the Type $A$ (Type $B$) solutions are singular along the negative (positive) $z$-axis. The 't Hooft magnetic fields of these solutions at spatial infinity correspond to the magnetic field of a positive one-half monopole located at the origin, $r=0$, and a semi-infinite Dirac string located on one-half of the $z$-axis which carries a magnetic flux of $\frac{2\pi}{g}$ going into the center of the sphere at infinity. Hence the net magnetic charge of the configuration is zero. The location of the Dirac string is always on the opposite side of the $z$-axis of its energy density concentration. 

When $\lambda=0$, the dimensionless total energy of the Type 1 one-half monopole solutions is 0.509 and the that of the Type 2 solutions is 0.525. Hence the total energies of these one-half monopole solutions seem to be higher than the dimensionless total energy of the BPS solution which is exactly $\frac{1}{2}$. The total energies  of these one-half monopole solutions are calculated for various strength of the Higgs field self-coupling constant and they are found to increase with $\lambda$. 

Recently we have constructed one-half monopole configurations which are axially symmetric and possess a Dirac string along one-half of the $z$-axis \cite{kn:9}. The magnetic fields are the usual magnetic field of a one-half monopole charge that is spherically symmetric and radial in direction. 

Other works on one-half monopoles in the Yang-Mills (YM) theory include the work of Harikumar et al. \cite{kn:10}. They demonstrated the existence of generic smooth YM potentials of one-half monopoles. However no exact or numerical solutions have been given. Exact one-half monopole axially symmetric solutions and one-half monopole mirror symmetric solutions with Dirac-like string have also been discussed in Ref. \cite{kn:11}.

We briefly review the SU(2) YMH field theory in the next section. The magnetic ansatz used in obtaining the one-half monopole solutions and some of its basic properties are discussed in section 3. We present the method used in obtaining the exact one-half monopole solutions at large distances in section 4 \cite{kn:9}. The numerical finite energy one-half magnetic monopole solutions and some of its properties are presented and discussed in section 5. We end with some comments in section 6.


\section{The SU(2) YMH Theory}

The SU(2) YMH Lagrangian in 3+1 dimensions with non vanishing Higgs potential is given by
\begin{equation}
{\cal L} = -\frac{1}{4}F^a_{\mu\nu} F^{a\mu\nu} - \frac{1}{2}D^\mu \Phi^a D_\mu \Phi^a - \frac{1}{4}\lambda(\Phi^a\Phi^a - \frac{\mu^2}{\lambda})^2. 
\label{eq.1}
\end{equation}

\noindent Here the Higgs field mass is $\mu$ and the strength of the Higgs potential is $\lambda$ which are constants. The vacuum expectation value of the Higgs field is $\xi=\mu/\sqrt{\lambda}$. The Lagrangian (\ref{eq.1}) is gauge invariant under the set of independent local SU(2) transformations at each space-time point.
The covariant derivative of the Higgs field and the gauge field strength tensor are given respectively by 
\begin{eqnarray}
D_{\mu}\Phi^{a} &=& \partial_{\mu} \Phi^{a} + g\epsilon^{abc} A^{b}_{\mu}\Phi^{c},\nonumber\\
F^a_{\mu\nu} &=& \partial_{\mu}A^a_\nu - \partial_{\nu}A^a_\mu + g\epsilon^{abc}A^b_{\mu}A^c_\nu,
\label{eq.2}
\end{eqnarray}
where $g$ is the gauge field coupling constant. The metric used is $g_{\mu\nu} = (- + + +)$. The SU(2) internal group indices $a, b, c = 1, 2, 3$  and the space-time indices are $\mu, \nu, \alpha = 0, 1, 2$, and $3$ in Minkowski space. The equations of motion that follow from the Lagrangian (\ref{eq.1}) are
\begin{eqnarray}
D^{\mu}F^a_{\mu\nu} &=& \partial^{\mu}F^a_{\mu\nu} + g\epsilon^{abc}A^{b\mu}F^c_{\mu\nu} = g\epsilon^{abc}\Phi^{b}D_{\nu}\Phi^c,\nonumber\\
D^{\mu}D_{\mu}\Phi^a &=& \lambda\Phi^a(\Phi^{b}\Phi^{b} - \xi^2).
\label{eq.3}
\end{eqnarray}
In the limit of vanishing $\mu$ and $\lambda$, the Higgs potential vanishes and self-dual solutions can be obtained by solving the first order partial differential Bogomol'nyi equation, 
\begin{equation}
B^a_i \pm D_i \Phi^a = 0, ~~~\mbox{where}~ B^a_i=-\frac{1}{2}\epsilon_{ijk}F^a_{jk}.
\label{eq.4}
\end{equation}

The electromagnetic field tensor proposed by 't Hooft \cite{kn:1} upon symmetry breaking is
\begin{eqnarray}
F_{\mu\nu} &=& \hat{\Phi}^a F^a_{\mu\nu} - \frac{1}{g}\epsilon^{abc}\hat{\Phi}^{a}D_{\mu}\hat{\Phi}^{b}D_{\nu}\hat{\Phi}^c,\nonumber\\
	&=& \partial_{\mu}A_\nu - \partial_{\nu}A_\mu - \frac{1}{g}\epsilon^{abc}\hat{\Phi}^{a}\partial_{\mu}\hat{\Phi}^{b}\partial_{\nu}\hat{\Phi}^c, ~~~\mbox{where}
\label{eq.5}\\
G_{\mu\nu} &=&  \partial_{\mu}A_\nu - \partial_{\nu}A_\mu, ~\mbox{and}~ H_{\mu\nu} = - \frac{1}{g}\epsilon^{abc}\hat{\Phi}^{a}\partial_{\mu}\hat{\Phi}^{b}\partial_{\nu}\hat{\Phi}^c,
\label{eq.6}
\end{eqnarray}

\noindent are the gauge part and the Higgs part of the electromagnetic field respectively. Here $A_\mu = \hat{\Phi}^{a}A^a_\mu$, the Higgs unit vector, $\hat{\Phi}^a = \Phi^a/|\Phi|$, and the Higgs field magnitude $|\Phi| = \sqrt{\Phi^{a}\Phi^{a}}$. 
Hence the decomposed magnetic field is
\begin{eqnarray}
B_i = -\frac{1}{2}\epsilon_{ijk}F_{jk}
      = B^G_i + B^H_i,
\label{eq.7}
\end{eqnarray}
where $B_i^G$ and $B_i^H$ are the gauge part and Higgs part of the magnetic field respectively. The net magnetic charge of the system is
\begin{eqnarray}
M = \frac{1}{4\pi} \int \partial^i B_i ~d^{3}x  = \frac{1}{4\pi} \oint d^{2}\sigma_{i}~B_i.
\label{eq.8}
\end{eqnarray}

The topological magnetic current, \cite{kn:12} 
\begin{equation*}
k_\mu = \frac{1}{8\pi}~\epsilon_{\mu\nu\rho\sigma}~\epsilon_{abc}~\partial^{\nu}\hat{\Phi}^{a}~\partial^{\rho}\hat{\Phi}^{b}~\partial^{\sigma}\hat{\Phi}^c,
\label{eq.9a}
\end{equation*}

\noindent is also the topological current density of the system. Hence the corresponding conserved topological magnetic charge is
\begin{eqnarray}
M_H & = & \frac{1}{g}\int d^{3}x~k_0 = \frac{1}{8\pi g}\int \epsilon_{ijk}\epsilon^{abc}\partial_{i}\left(\hat{\Phi}^{a}\partial_{j}\hat{\Phi}^{b}\partial_{k}\hat{\Phi}^{c}\right)d^{3}x \nonumber\\
& = & \frac{1}{8\pi}\oint d^{2}\sigma_{i}\left(\frac{1}{g}\epsilon_{ijk}\epsilon^{abc}\hat{\Phi}^{a}\partial_{j}\hat{\Phi}^{b}\partial_{k}\hat{\Phi}^{c}\right)\nonumber\\
& = & \frac{1}{4\pi} \oint d^{2}\sigma_{i}~B_i^H. 
\label{eq.9}
\end{eqnarray}
\noindent The magnetic charge $M_H$ is the total magnetic charge of the system if and only if the gauge field is non singular \cite{kn:13}. If the gauge field is singular and carries Dirac string monopoles $M_G$, then the total magnetic charge of the system is
\begin{eqnarray}
M &=& M_G + M_H, ~~\mbox{where}\nonumber\\
M_G & = & -\frac{1}{8\pi}\oint d^{2}\sigma_{i}\epsilon_{ijk}\left(\partial_j A_k - \partial_k A_j\right)\nonumber\\
& = & \frac{1}{4\pi} \oint d^{2}\sigma_{i}~B_i^G. 
\label{eq.10}
\end{eqnarray}

In the electrically neutral BPS limit when the Higgs potential vanishes, the energy is a minimum, \cite{kn:14}
\begin{eqnarray}
E_{min} &=& \mp\int\partial_i(B^a_i\Phi^a)~d^3 x + \int\frac{1}{2}(B^a_i \pm D_i\Phi^a)^2~d^3 x\nonumber\\
&=& \mp\int\partial_i(B^a_i\Phi^a)~d^3 x = \frac{4\pi\xi}{g} M_H,
\label{eq.11}
\end{eqnarray}
when the vacuum expectation value of the Higgs field is non zero. Hence the dimensionless minimum total energy is $M_H$.

For a non BPS solution, its energy must be greater than that given by Eq. (\ref{eq.11}). The dimensionless value is given by
\begin{eqnarray}
E=\frac{g}{8\pi\xi}\int \{B^a_iB^a_i +  D_i\Phi^aD_i\Phi^a + \frac{\lambda}{2}(\Phi^a\Phi^a-\xi^2)^2\}~d^3x \geq M_H.
\label{eq.12}
\end{eqnarray}

The gauge potential $A^a_\mu$ and the Higgs field $\Phi^a$ can be gauge transformed by local SU(2) gauge transformations which can be written in the $2\times 2$ matrix form, \cite{kn:14}
\begin{eqnarray}
\omega(x)=\exp\left(\frac{i}{2}\sigma_a\hat{n}^a(x) f(x)\right) = \cos\left(\frac{1}{2}f(x)\right)+i\sigma_a\hat{n}^a(x)\sin\left(\frac{1}{2}f(x)\right), 
\label{eq.13}
\end{eqnarray}
where $\hat{n}^a(x)$ is a unit vector and $\sigma_a$ is the Pauli matrix. The transformed gauge potential and Higgs field then takes the form,
\begin{eqnarray}
A^{\prime a}_\mu &=& \cos f A^a_\mu + \sin f \epsilon_{abc}A^b_\mu\hat{n}^c + 2\sin^2\frac{f}{2}\hat{n}^a(\hat{n}_b A^b_\mu) \nonumber\\
&+& \left\{\hat{n}^a\partial_\mu f + \sin f \partial_\mu \hat{n}^a + 2\sin^2\frac{f}{2}\epsilon_{abc}(\partial_\mu \hat{n}^b) \hat{n}^c\right\}
\label{eq.14}\\
\Phi^{\prime a}&=&\cos f \Phi^a + \sin f \epsilon_{abc}\Phi^b \hat{n}^c + 2\sin^2 \frac{f}{2}\hat{n}^a (\hat{n}^b \sigma_b).
\label{eq.15}
\end{eqnarray}


\section{The Magnetic Ansatz}
The magnetic ansatz  \cite{kn:8} is given by
\begin{eqnarray}
gA_i^a &=&  - \frac{1}{r}\psi_1(r, \theta) \hat{n}^{a}_\phi\hat{\theta}_i + \frac{1}{r\sin\theta}P_1(r, \theta)\hat{n}^{a}_\theta\hat{\phi}_i
+ \frac{1}{r}R_1(r, \theta)\hat{n}^{a}_\phi\hat{r}_i - \frac{1}{r\sin\theta}P_2(r, \theta)\hat{n}^{a}_r\hat{\phi}_i, \nonumber\\
gA^a_0 &=& 0, ~~~g\Phi^a = \Phi_1(r, \theta)~\hat{n}^a_r + \Phi_2(r, \theta)\hat{n}^a_\theta,
\label{eq.16}
\end{eqnarray}
\noindent where $P_1(r, \theta)=\sin\theta~\psi_2(r, \theta)$ and $P_2(r, \theta)=\sin\theta~R_2(r, \theta)$. The spatial spherical coordinate orthonormal unit vectors are
\begin{eqnarray}
\hat{r}_i &=& \sin\theta ~\cos \phi ~\delta_{i1} + \sin\theta ~\sin \phi ~\delta_{i2} + \cos\theta~\delta_{i3}, \nonumber\\
\hat{\theta}_i &=& \cos\theta ~\cos \phi ~\delta_{i1} + \cos\theta ~\sin \phi ~\delta_{i2} - \sin\theta ~\delta_{i3}, \nonumber\\
\hat{\phi}_i &=& -\sin \phi ~\delta_{i1} + \cos \phi ~\delta_{i2},
\label{eq.17}
\end{eqnarray}
and the isospin coordinate orthonormal unit vectors are 
\begin{eqnarray}
\hat{n}_r^a &=& \sin \theta ~\cos n\phi ~\delta_{1}^a + \sin \theta ~\sin n\phi ~\delta_{2}^a + \cos \theta~\delta_{3}^a,\nonumber\\
\hat{n}_\theta^a &=& \cos \theta ~\cos n\phi ~\delta_{1}^a + \cos \theta ~\sin n\phi ~\delta_{2}^a - \sin \theta ~\delta_{3}^a,\nonumber\\
\hat{n}_\phi^a &=& -\sin n\phi ~\delta_{1}^a + \cos n\phi ~\delta_{2}^a; ~~~\mbox{where}~~n\geq 1.
\label{eq.18}
\end{eqnarray}
The $\phi$-winding number $n$ is a natural number. In our work here on one-half monopole we take $n=1$.
The magnetic ansatz (\ref{eq.16}) is form invariant under the gauge transformation
\begin{eqnarray}
\omega = \exp\left(\frac{i}{2}\sigma^a\hat{n}^a_{\phi} f(r,\theta)\right), ~~~\sigma^a = ~\mbox{Pauli matrices}
\label{eq.19}
\end{eqnarray}
and the transformed gauge potential and Higgs field take the form,
\begin{eqnarray}
gA_i^{\prime a} &=&  - \frac{1}{r}\{\psi_1 - \partial_\theta f\}\hat{n}^{a}_\phi\hat{\theta}_i \nonumber\\
&+& \frac{1}{r\sin\theta}\left\{P_1\cos f + P_2\sin f + n[\sin\theta-\sin(f+\theta)]\right\}\hat{n}^{a}_\theta\hat{\phi}_i\nonumber\\
&+& \frac{1}{r}\{R_1+r\partial_r f\}\hat{n}^{a}_\phi\hat{r}_i \nonumber\\
&-& \frac{1}{r\sin\theta}\left\{P_2\cos f - P_1\sin f - n[\cos\theta - \cos(f+\theta)]\right\}\hat{n}^{a}_r\hat{\phi}_i, \nonumber\\
gA^{\prime a}_0 &=& 0,  \nonumber\\
g\Phi^{\prime a} &=& (\Phi_1\cos f+\Phi_2\sin f)~\hat{n}^a_r + (\Phi_2\cos f-\Phi_1\sin f)\hat{n}^a_\theta.
\label{eq.20}
\end{eqnarray}

The general Higgs fields in the spherical and the rectangular coordinate systems are
\begin{eqnarray}
g\Phi^a &=& \Phi_1(x)~\hat{n}^a_r + \Phi_2(x)\hat{n}^a_\theta + \Phi_3(x)\hat{n}^a_\phi\nonumber\\
&=& \tilde{\Phi}_1(x) ~\delta^{a1} + \tilde{\Phi}_2(x) ~\delta^{a2} + \tilde{\Phi}_3(x) ~\delta^{a3},
\label{eq.21}
\end{eqnarray}
respectively, where
\begin{eqnarray}
\tilde{\Phi}_1 &=& \sin\theta \cos n\phi ~\Phi_1 + \cos\theta \cos n\phi ~\Phi_2 - \sin n\phi ~\Phi_3
= |\Phi|\sin\alpha \cos\beta\nonumber\\
\tilde{\Phi}_2 &=& \sin\theta \sin n\phi ~\Phi_1 + \cos\theta \sin n\phi ~\Phi_2 + \cos n\phi ~\Phi_3
= |\Phi|\sin\alpha \sin\beta\nonumber\\
\tilde{\Phi}_3 &=& \cos\theta ~\Phi_1 - \sin\theta ~\Phi_2 = |\Phi|\cos\alpha.
\label{eq.22}
\end{eqnarray}
The axially symmetric Higgs unit vector in the rectangular coordinate system is
\begin{eqnarray}
\hat{\Phi}^a &=& \sin\alpha \cos\beta ~\delta^{a1} + \sin\alpha \sin\beta ~\delta^{a2} + \cos\alpha ~\delta^{a3}, 
\label{eq.23}\\
\cos\alpha &=& g(r,\theta)\cos\theta - h(r,\theta)\sin\theta,~~~\beta=n\phi,
\label{eq.24}\\
g(r,\theta) &=& \frac{\Phi_1}{|\Phi|}, ~~~h(r,\theta) = \frac{\Phi_2}{|\Phi|}.\nonumber
\end{eqnarray}
From Eq. (\ref{eq.23}) and (\ref{eq.24}), the Higgs field (\ref{eq.16}) and the gauge transformed Higgs field (\ref{eq.20}) can be written in terms of $\alpha$ as
\begin{eqnarray}
\Phi^a &=& |\Phi(r,\theta)|(\cos(\alpha-\theta)~\hat{n}^a_r + \sin(\alpha-\theta)\hat{n}^a_\theta),\nonumber\\
\Phi^{\prime a} &=& |\Phi(r,\theta)|(\cos(\alpha^\prime-\theta)~\hat{n}^a_r + \sin(\alpha^\prime-\theta)\hat{n}^a_\theta),
\label{eq.25}
\end{eqnarray}
where $\alpha^\prime=\alpha-f$.

Similarly the Higgs part of the 't Hooft magnetic field (\ref{eq.7}) can be reduced to
\begin{eqnarray}
gB_i^H = -n\epsilon_{ijk} \partial^j\cos\alpha\partial^k\phi.
\label{eq.26}
\end{eqnarray}
The gauge part of the magnetic field (\ref{eq.7}) can be written in similar form
\begin{eqnarray}
gB^G_i &=& -n\epsilon_{ijk}\partial_j\cos\kappa ~\partial_k \phi,
 \label{eq.27}\\
\mbox{where},~~\cos\kappa &=& \frac{1}{n}\left(h(r,\theta)P_1 - g(r,\theta)P_2\right).\nonumber
\end{eqnarray}

\noindent Hence the 't Hooft's magnetic field which is the sum of the Higgs part (\ref{eq.26}) and the gauge part (\ref{eq.27}) is given by
\begin{eqnarray}
gB_i = -n\epsilon_{ijk}\partial_j(\cos\alpha + \cos\kappa)~\partial_k \phi = -\epsilon_{ijk}\partial_j{\cal A}_k,
\label{eq.28}
\end{eqnarray}
where ${\cal A}_i$ is the 't Hooft's gauge potential.
The magnetic field lines of the configuration can be plotted by drawing the contour lines of $(\cos\alpha + \cos\kappa) = $ constant on the vertical plane $\phi=0$. The orientation of the magnetic field can also be plotted by using the vector plot of the magnetic field unit vector,
\begin{eqnarray}
\hat{B}_i = \frac{-\partial_\theta (\cos\alpha + \cos\kappa)\hat{r}_i + r \partial_r (\cos\alpha + \cos\kappa)\hat{\theta}_i}{\sqrt{[r \partial_r (\cos\alpha + \cos\kappa)]^2 + [\partial_\theta (\cos\alpha + \cos\kappa)]^2}}.
\label{eq.29}
\end{eqnarray}

At spatial infinity in the Higgs vacuum, all the non-Abelian components of the gauge potential vanish and the non-Abelian electromagnetic field tends to 
\begin{eqnarray}
\left.F^a_{\mu\nu}\right|_{r\rightarrow\infty} &=& \{\partial_\mu A_\nu - \partial_\nu A_\mu - \frac{1}{g}\epsilon^{cde}\hat{\Phi}^c\partial_\mu \hat{\Phi}^d \partial_\nu \hat{\Phi}^e\}\hat{\Phi}^a \nonumber\\
						&=& F_{\mu\nu}\hat{\Phi}^a, 
\label{eq.30}
\end{eqnarray}
where $F_{\mu\nu}$ is just the 't Hooft electromagnetic field. However there is no unique way of representing the Abelian electromagnetic field in the region of the monopole outside the Higgs vacuum at finite values of $r$ \cite{kn:15}. One proposal was given by 't Hooft as in Eq. (\ref{eq.5}). 
The 't Hooft's magnetic field has the special property, that the magnetic charge density vanishes, that is $\partial^i B_i=0$, when $|\Phi|\not=0$. However when $|\Phi|=0$, $\partial^i B_i\not=0$ and the magnetic charges are located at these points. Hence with 't Hooft's definition of the electromagnetic field, the magnetic charges are discrete and reside at the point zeros of the Higgs field. Since this definition gives a discrete magnetic charge at a particular point, there is no magnetic charge distribution over space. 

Another proposal which is less singular was given by Bogomol'nyi \cite{kn:3} and Faddeev \cite{kn:16},
\begin{eqnarray}
{\cal B}_i = B_i^a \left(\frac{\Phi^a}{\xi}\right), ~~~{\cal E}_i = E_i^a \left(\frac{\Phi^a}{\xi}\right).
\label{eq.31}
\end{eqnarray}
With this definition of the electromagnetic field (\ref{eq.31}), there will be a magnetic charge density distribution contributed by the non-Abelian components of the gauge field in the finite $r$ region. A 3D surface plot of the magnetic charge density distribution will be useful in determining the sign of the magnetic charge of the solutions.
In the Higgs vacuum at spatial infinity, both definitions of the electromagnetic field (\ref{eq.5}) and (\ref{eq.31}) become similar.


\section{The Exact Asymptotic Solutions}

The gauge transformation (\ref{eq.19}),  
when applied on to the solution,
\begin{eqnarray}
gA^a_0, ~~~gA^a_i = A(r,\theta)\delta^a_3\hat{\phi}_i, ~~~g\Phi^a=\xi\delta^a_3,\\
\mbox{where}~~ A(r,\theta)=\frac{a\cos\theta + b -1}{r\sin\theta}, ~~a, b = \mbox{constants}. \nonumber
\label{eq.32}
\end{eqnarray}
will rotate the isospin direction, $\delta^a_3$, into the magnetic anstaz (\ref{eq.16}) and the transformed gauge potentials and Higgs field then take the form,
\begin{eqnarray}
gA_i^{\prime a} &=&  - \frac{1}{r}\{-\partial_\theta f(r,\theta)\}\hat{n}^{a}_\phi\hat{\theta}_i + \frac{1}{r}\{\partial_r f(r,\theta)\}\hat{n}^{a}_\phi\hat{r}_i \nonumber\\
&+& \frac{n}{r\sin\theta}\left\{\sin\theta-\sin(f(r,\theta)+\theta)[1+rA(r,\theta)\sin\theta]\right\}\hat{n}^{a}_\theta\hat{\phi}_i\nonumber\\
&-& \frac{n}{r\sin\theta}\left\{\cos\theta - \cos(f(r,\theta)+\theta)[1+rA(r,\theta)\sin\theta]\right\}\hat{n}^{a}_r\hat{\phi}_i, \nonumber\\
gA^{\prime a}_0 &=& 0,  \nonumber\\
g\Phi^{\prime a} &=& \xi \{\cos(f(r,\theta)+\theta)~\hat{n}^a_r - \sin(f(r,\theta)+\theta)\hat{n}^a_\theta\}.
\label{eq.33}
\end{eqnarray}

\noindent We also note that the angle $\alpha=-f(r,\theta)$. Hence
\begin{eqnarray}
\cos\alpha=\cos f(r,\theta), ~~~\cos\kappa = -\cos f(r,\theta) + \left\{1+r\sin\theta A(r,\theta)\right\},
\label{eq.34}
\end{eqnarray}
and the 't Hooft gauge potential and net magnetic field of the monopole configuration are given by
\begin{eqnarray}
{\cal A}_i &=& \left\{\frac{a\cos\theta+b}{r\sin\theta}\right\}\hat{\phi}_i, \nonumber\\
gB_i &=& -\epsilon_{ijk}\partial_j\{1+r\sin\theta A(r,\theta)\}\partial_k\phi = \frac{na}{r^2}\hat{r}_i,
\label{eq.35}
\end{eqnarray}
respectively.
Hence one-half magnetic monopole solutions of the magnetic ansatz (\ref{eq.16}) can be obtained when $n=1$ and $a=1/2$.

The Type $A$1 solution is obtained by letting $f(r,\theta) = -\frac{1}{2}\theta$, $a=\frac{1}{2}$, $b=\frac{1}{2}$, and $n=1$. The profile functions of the gauge potentials are given by,
\begin{eqnarray}
\psi_1&=& \frac{1}{2}, ~~~P_1 = \sin\theta - \frac{1}{2}\sin \frac{1}{2}\theta (1+\cos\theta), \nonumber\\
R_1&=&0, ~~~P_2=\cos\theta - \frac{1}{2}\cos \frac{1}{2}\theta(1+\cos\theta), \nonumber\\
\Phi_1&=& \xi \cos \frac{1}{2}\theta,~~~\Phi_2=-\xi \sin \frac{1}{2}\theta.
\label{eq.36}
\end{eqnarray}

\noindent The Type $A$2 solution is obtained by letting $f(r,\theta) = \frac{1}{2}\theta$, $a=\frac{1}{2}$, $b=\frac{1}{2}$, and $n=1$. The profile functions of the gauge potentials are given by,
\begin{eqnarray}
\psi_1&=& -\frac{1}{2}, ~~~P_1 = \sin\theta - \frac{1}{2}\sin \frac{3}{2}\theta (1+\cos\theta), \nonumber\\
R_1&=&0, ~~~P_2=\cos\theta - \frac{1}{2}\cos \frac{3}{2}\theta(1+\cos\theta), \nonumber\\
\Phi_1&=& \xi \cos \frac{3}{2}\theta,~~~\Phi_2=-\xi \sin \frac{3}{2}\theta.
\label{eq.37}
\end{eqnarray}

\noindent Both Type $A$ solutions possess a magnetic charge of $\frac{1}{2g}$ at $r=0$ and 't Hooft's gauge potential, ${\cal A}_i = \left\{\frac{\cos\theta + 1}{2r\sin\theta}\right\}\hat{\phi}_i$, which is singular along the positive $z$-axis.

The Type $B$1 solution is obtained by letting $f(r,\theta) = -\frac{1}{2}\theta-\frac{\pi}{2}$, $a=\frac{1}{2}$, $b=-\frac{1}{2}$, and $n=1$. The profile functions of the gauge potentials are given by,
\begin{eqnarray}
\psi_1&=& \frac{1}{2}, ~~~P_1 = \sin\theta - \frac{1}{2}\cos \frac{1}{2}\theta (1-\cos\theta), \nonumber\\
R_1&=&0, ~~~P_2=\cos\theta + \frac{1}{2}\sin \frac{1}{2}\theta(1-\cos\theta), \nonumber\\
\Phi_1&=& \xi \sin \frac{1}{2}\theta,~~~\Phi_2=\xi \cos \frac{1}{2}\theta.
\label{eq.38}
\end{eqnarray}

\noindent The Type $B$2 solution is obtained by letting $f(r,\theta) = \frac{1}{2}\theta + \frac{\pi}{2}$, $a=\frac{1}{2}$, $b=-\frac{1}{2}$, and $n=1$. The profile functions of the gauge potentials are given by,
\begin{eqnarray}
\psi_1&=& -\frac{1}{2}, ~~~P_1 = \sin\theta + \frac{1}{2}\cos \frac{3}{2}\theta (1-\cos\theta), \nonumber\\
R_1&=&0, ~~~P_2=\cos\theta - \frac{1}{2}\sin \frac{3}{2}\theta(1-\cos\theta), \nonumber\\
\Phi_1&=& -\xi \sin \frac{3}{2}\theta,~~~\Phi_2=-\xi \cos \frac{3}{2}\theta.
\label{eq.39}
\end{eqnarray}

\noindent Both Type $B$ solutions possess a magnetic charge of $\frac{1}{2g}$ at $r=0$ and 't Hooft's gauge potential, ${\cal A}_i = \left\{\frac{\cos\theta - 1}{2r\sin\theta}\right\}\hat{\phi}_i$, which is singular along the negative $z$-axis. 

The magnetic field $B^a_i$ of the above four solutions are given by
\begin{eqnarray}
gB^a_i = \frac{\hat{r}_i}{2 r^2}\hat{\Phi}^a,
\label{eq.40}
\end{eqnarray}
when the semi-infinite Dirac string along one-half of the $z$-axis is ignored.
Thus the four solutions possess common magnetic field of a one-half monopole located at the origin, $r=0$, that are spherically symmetric, radial in direction and pointing in the Higgs field direction in isospin space. 

Similar to the gauge potentials, the Higgs magnetic fields of the Type $A$ solutions are singular along the negative $z$-axis, whereas the Higgs magnetic fields of the Type $B$ solutions are singular along the positive $z$-axis and they are given by
\begin{eqnarray}
gB^H_i = \frac{\frac{1}{2}\sin(\frac{1}{2}\theta)}{\sin\theta}\frac{\hat{r}_i}{r^2}, ~~~~~\mbox{Type}~A ~~\mbox{solutions}
\label{eq.41}\\
gB^H_i = \frac{\frac{1}{2}\cos(\frac{1}{2}\theta)}{\sin\theta}\frac{\hat{r}_i}{r^2}, ~~~~~\mbox{Type}~B ~~\mbox{solutions}.
\label{eq.42}
\end{eqnarray}

\noindent The net 't Hooft's magnetic fields for the Type $A$ and Type $B$ solutions are however given by 
\begin{eqnarray}
gB_i = \frac{\hat{r}_i}{2 r^2} - 2\pi\delta(x_1)\delta(x_2)\theta(x_3)\delta^3_i, ~~~~~\mbox{Type}~A ~~\mbox{solutions},
\label{eq.43}\\
gB_i = \frac{\hat{r}_i}{2 r^2} + 2\pi\delta(x_1)\delta(x_2)\theta(-x_3)\delta^3_i, ~~~~~\mbox{Type}~B ~~\mbox{solutions},
\label{eq.44}
\end{eqnarray}
when the Dirac string in the gauge potentials, ${\cal A}_i$, is taken into consideration \cite{kn:17}. These one-half monopole solutions therefore carry a positive one-half magnetic monopole at the origin and a semi-infinite Dirac string of flux, $\frac{2\pi}{g}$, going into the origin. Hence the net magnetic charge of the configuration is zero.


\section{The Numerical Solutions}
\subsection{The Numerical Calculations}
\label{subsection 5.1}

The numerical one-half monopole solutions can be constructed by making use of the exact one-half magnetic monopole solutions (\ref{eq.36}) - (\ref{eq.39}) as asymptotic solutions at large distances and by fixing the boundary conditions for the profile functions (\ref{eq.16}) along the $z$-axis and near $r=0$. Hence, we are able to obtain four types of finite energy numerical one-half magnetic monopole solutions which we will name as the numerical Type $A$1, $A$2, $B$1, and $B$2 one-half monopole solutions respectively. 
Since the function $R_2(r,\theta)$ is singular along one side of the $z$-axis at infinity, we choose to perform our numerical analysis with the functions,
\begin{eqnarray}
P_1(r,\theta)=\psi_2(r,\theta)\sin\theta, ~~P_2(r,\theta)=R_2(r,\theta)\sin\theta.
\label{eq.45}
\end{eqnarray}

Near the origin, $r=0$, we have the common trivial vacuum solution for the four solutions. The asymptotic solutions and boundary conditions at small distances that will give rise to finite energy solutions are
\begin{eqnarray}
\psi_1=P_1=R_1=P_2=0,~~~\Phi_1=\xi_0\cos\theta, ~~~\Phi_2=-\xi_0\sin\theta,
\label{eq.46}\\
\sin\theta\Phi_1(0,\theta)+\cos\theta\Phi_2(0,\theta)=0,\nonumber\\
\partial_r(\cos\theta\Phi_1(r,\theta)-\sin\theta\Phi_2(r,\theta))|_{r=0}=0.
\label{eq.47}
\end{eqnarray}

\noindent The boundary conditions imposed along the positive $z$-axis for the profile functions (\ref{eq.16}) of the Type $A$ solutions are
\begin{eqnarray}
\partial_\theta \Phi_1(r,\theta)|_{\theta=0} = 0, ~~\Phi_2(r,0)=0,~~ \partial_\theta \psi_1(r,\theta)|_{\theta=0} = 0,\nonumber\\
R_1(r,0)=0, ~~P_1(r,0)=0, ~~P_2(r,0)=0.
\label{eq.48}
\end{eqnarray}
Along the negative $z$-axis, the boundary conditions imposed are
\begin{eqnarray}
\Phi_1(r,\pi)=0,~~ \partial_\theta \Phi_2(r,\theta)|_{\theta=\pi} = 0, ~~\partial_\theta \psi_1(r,\theta)|_{\theta=\pi} = 0,\nonumber\\
R_1(r,\pi)=0, ~~P_1(r,\pi)=0, ~~\partial_\theta P_2(r,\theta)|_{\theta=\pi}=0.
\label{eq.49}
\end{eqnarray}

\noindent The boundary conditions imposed along the positive $z$-axis for the profile functions (\ref{eq.16}) of the Type $B$ solutions are
\begin{eqnarray}
\Phi_1(r,0)=0, ~~\partial_\theta \Phi_2(r,\theta)|_{\theta=0} = 0,~~ \partial_\theta \psi_1(r,\theta)|_{\theta=0} = 0,\nonumber\\
R_1(r,0)=0, ~~P_1(r,0)=0, ~~\partial_\theta P_2(r,\theta)|_{\theta=0}=0,
\label{eq.50}
\end{eqnarray}
and along the negative $z$-axis, the boundary conditions imposed are
\begin{eqnarray}
\partial_\theta \Phi_1(r,\theta)|_{\theta=\pi} = 0,~~ \Phi_2(r,\pi)=0, ~~\partial_\theta \psi_1(r,\theta)|_{\theta=\pi} = 0,\nonumber\\
R_1(r,\pi)=0, ~~P_1(r,\pi)=0, ~~P_2(r,\pi)=0.
\label{eq.51}
\end{eqnarray}

\noindent The numerical Type $A$1, $A$2, $B$2, and $B$2 solutions connecting the asymptotic solutions (\ref{eq.36}) - (\ref{eq.39}) respectively at large distances to the trivial vacuum solution (\ref{eq.46}) at small distances and subjected to the boundary conditions (\ref{eq.47}) - (\ref{eq.49}) for the Type $A$ solutions and the boundary conditions (\ref{eq.47}),  (\ref{eq.50}) - (\ref{eq.51}) for the Type $B$ solutions together with the gauge fixing condition \cite{kn:8}
\begin{equation}
r\partial_rR_1-\partial_\theta \psi_1=0,
\label{eq.52}
\end{equation}
are solved using the Maple 12 and MatLab R2009a softwares \cite{kn:18}. 

The second order equations of motion (\ref{eq.3}) which are reduced to six partial differential equations with the ansatz (\ref{eq.16}) are then transformed into a system of nonlinear equations using the finite difference approximation. The numerical calculations are performed with the system of nonlinear equations been discretized on a non-equidistant grid of size $90\times80$ covering the integration regions $0\leq \bar{x} \leq 1$ and $0\leq \theta \leq \pi$. Here $\bar{x}=\frac{r}{r+1}$ is the finite interval compactified coordinate. The partial derivative with respect to the radial coordinate is then replaced accordingly by ~$\partial_r \rightarrow (1-\bar{x})^2 \partial_{\bar{x}}$~ and ~$\frac{\partial^2}{\partial r^2} \rightarrow (1-\bar{x})^4\frac{\partial^2}{\partial \bar{x}^2} - 2(1-\bar{x})^3\frac{\partial}{\partial \bar{x}}$~. We first used Maple to find the Jacobian sparsity pattern for the system of nonlinear equations.  After that we provide this information to Matlab to run the numerical computation. The system of nonlinear equations are then solved numerically using the trust-region-reflective algorithm by providing the solver with good initial guess.
The second order equations of motion Eq. (\ref{eq.3}) are solved when the $\phi$-winding number $n=1$, with nonzero expectation value $\xi=1$, gauge coupling constant $g=1$, and when the Higgs self-coupling constant $0\leq\lambda\leq12$.
In our calculations, there are two kinds of error, the first inherited from the finite difference approximation of the functions, which is of the order of $10^{-4}$. The second comes from the linearization of the nonlinear equations for MATLAB to solve numerically; the remainder or approximation error is of the order of $10^{-6}$.
Hence the overall error estimate is $10^{-4}$.

\begin{figure}[tbh]
	\centering
	\hskip-0.1in
	 \includegraphics[width=6.0in,height=6.0in]{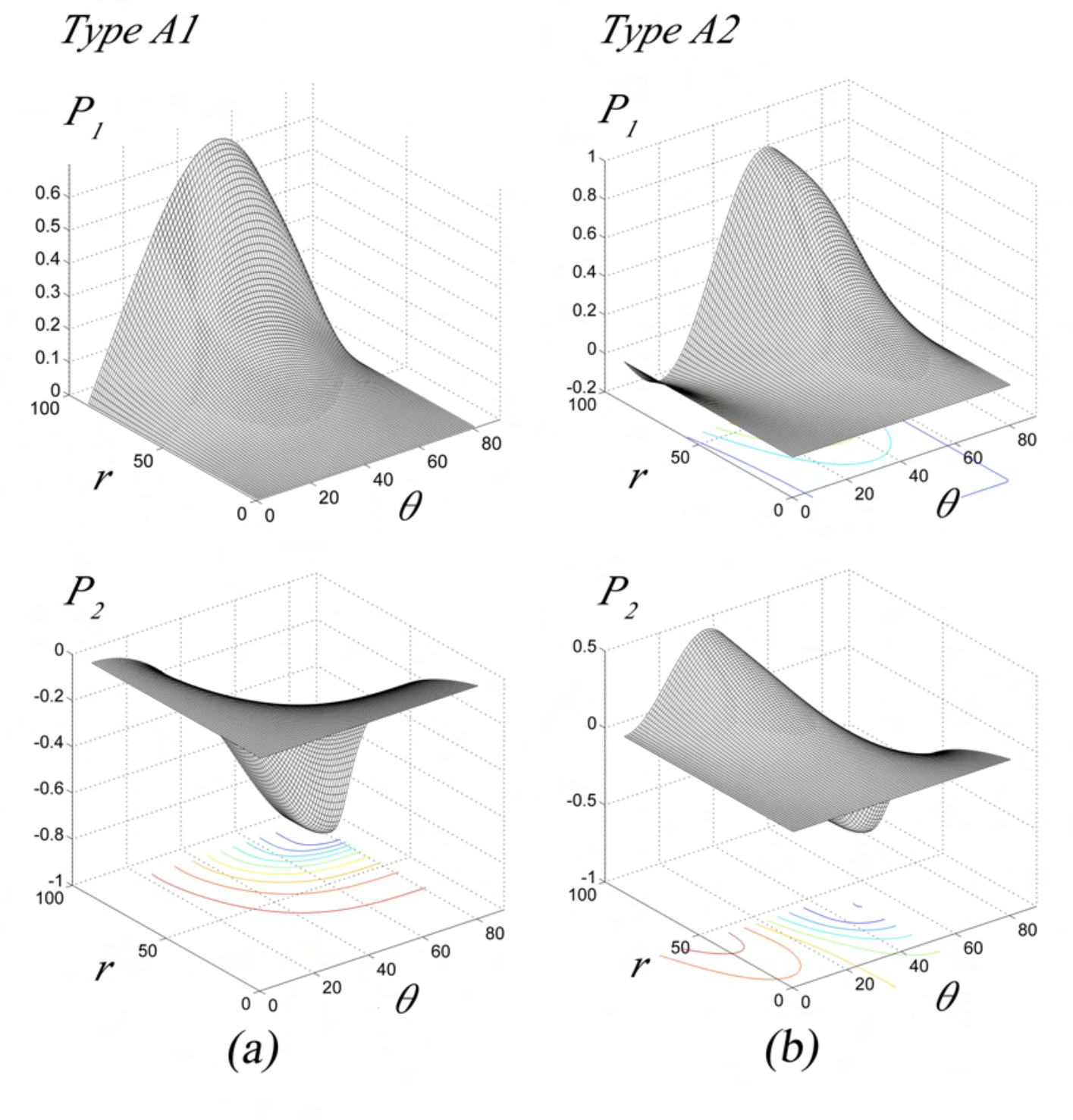} 
	\caption{The 3D profile function graphs $P_1(r,\theta)$ and $P_2(r,\theta)$ versus $r$ and $\theta$ in rectangular coordinate frame of grid size 90 x 80 of the (a) $A$1 and (b) $A$2 one-half monopole solutions when $\lambda=\xi=1$.}
	\label{fig.1}
\end{figure}

\begin{figure}[tbh]
	\centering
	\hskip0in
	 \includegraphics[width=6.0in,height=6.0in]{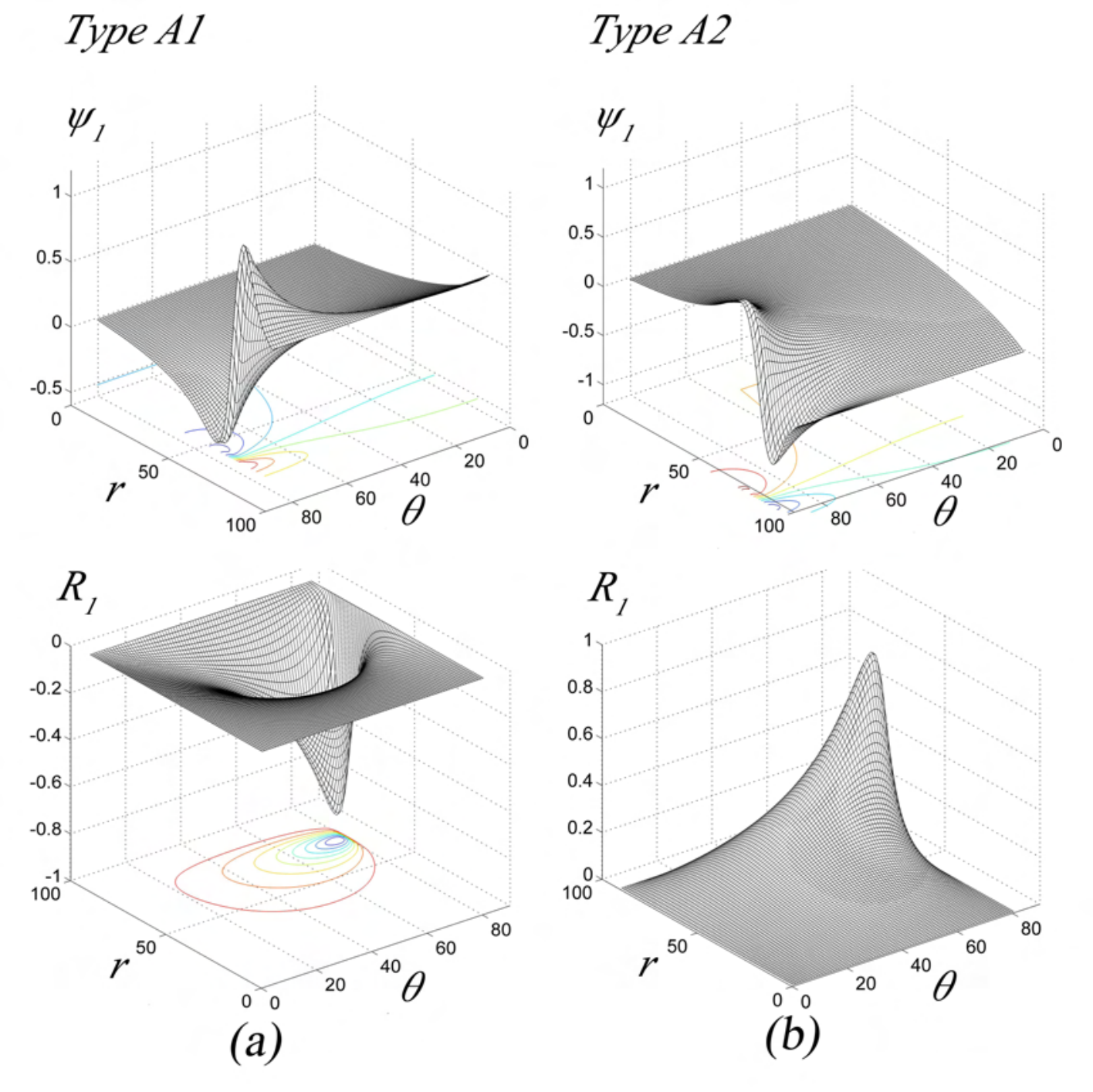} 
	\caption{The 3D profile function graphs $\psi_1(r,\theta)$ and $R_1(r,\theta)$ versus $r$ and $\theta$ in rectangular coordinate frame of grid size 90 x 80 of the (a) $A$1 and (b) $A$2 one-half monopole solutions when $\lambda=\xi=1$.}
	\label{fig.2}
\end{figure}

\begin{figure}[tbh]
	\centering
	\hskip-0.1in
	 \includegraphics[width=6.0in,height=6.0in]{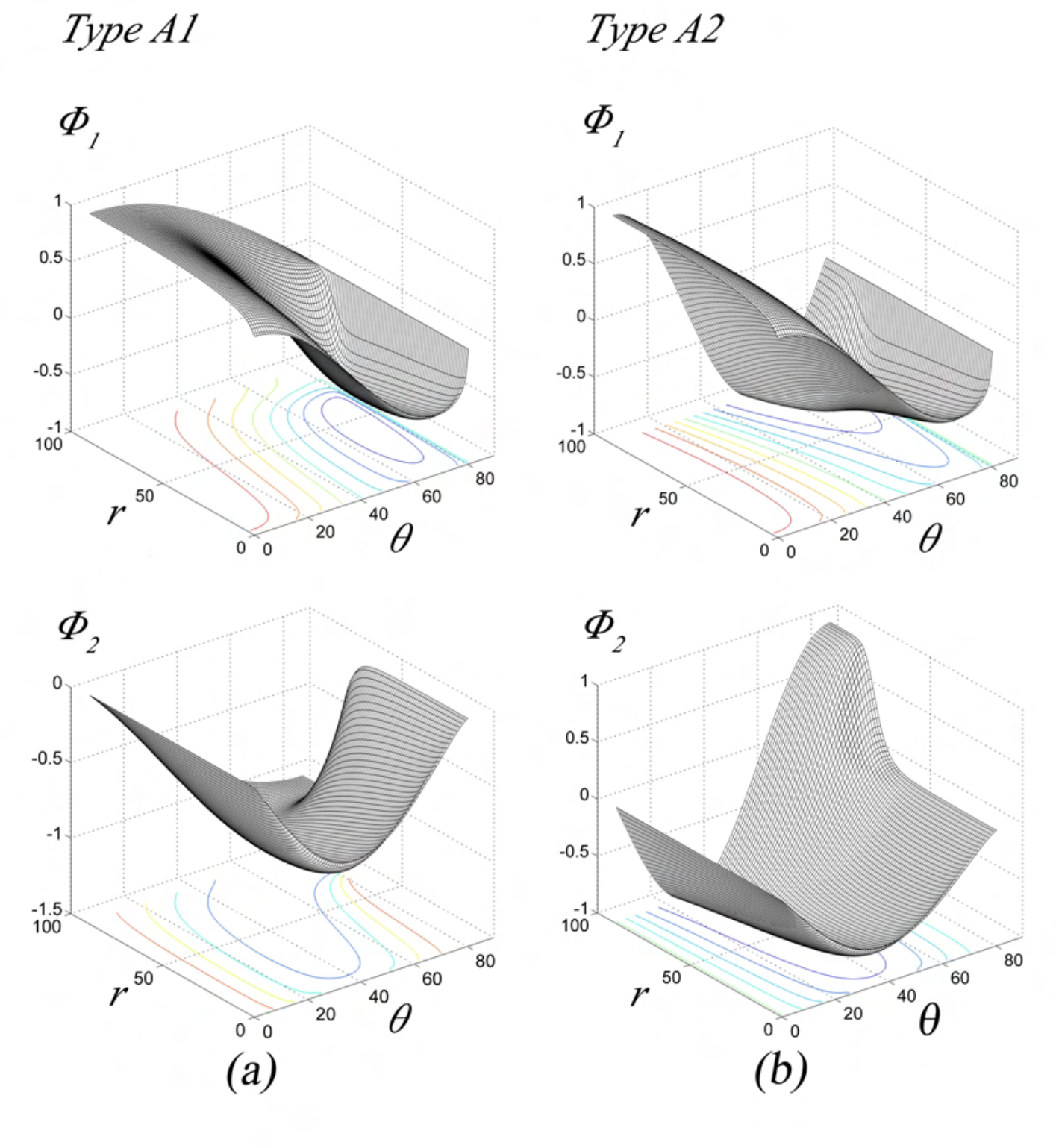} 
	\caption{The 3D profile function graphs $\Phi_1(r,\theta)$ and $\Phi_2(r,\theta)$ versus $r$ and $\theta$ in rectangular coordinate frame of grid size 90 x 80 of the (a) $A$1 and (b) $A$2 one-half monopole solutions when $\lambda=\xi=1$.}
	\label{fig.3}
\end{figure}

The numerical solutions obtained for the four one-half monopole solutions, $\psi_1, P_1, R_1, P_2, \Phi_1$, and $\Phi_2$, are all regular functions of $r$ and $\theta$. However the function $R_2=\frac{P_2}{\sin\theta}$ possesses a string singularity along the negative (positive) $z$-axis for the Type $A$ (Type $B$) solutions. 
Our results show that the Type $B$1 and Type $B$2 solutions are exactly $180^o$ rotation of the $z$-axis about the origin, $r=0$, of the Type $A$1 and Type $A$2 solutions respectively. The 3D profile function graphs versus $r$ and $\theta$ in rectangular coordinate frame of grid size 90 x 80 are shown for the Type $A$1 and Type $A$2 solutions, $P_1(r,\theta)$ and $P_2(r,\theta)$ in Figure \ref{fig.1}, $\psi_1(r,\theta)$ and $R_1(r,\theta)$ in Figure \ref{fig.2}, and $\Phi_1(r,\theta)$ and $\Phi_1(r,\theta)$ in Figure \ref{fig.3}. 


\subsection{The Magnetic Dipole Moment and Magnetic Charge}
\label{subsection 5.2}

\begin{figure}[tbh]
	\centering
	\hskip-0.1in
	 \includegraphics[width=5.8in,height=5.5in]{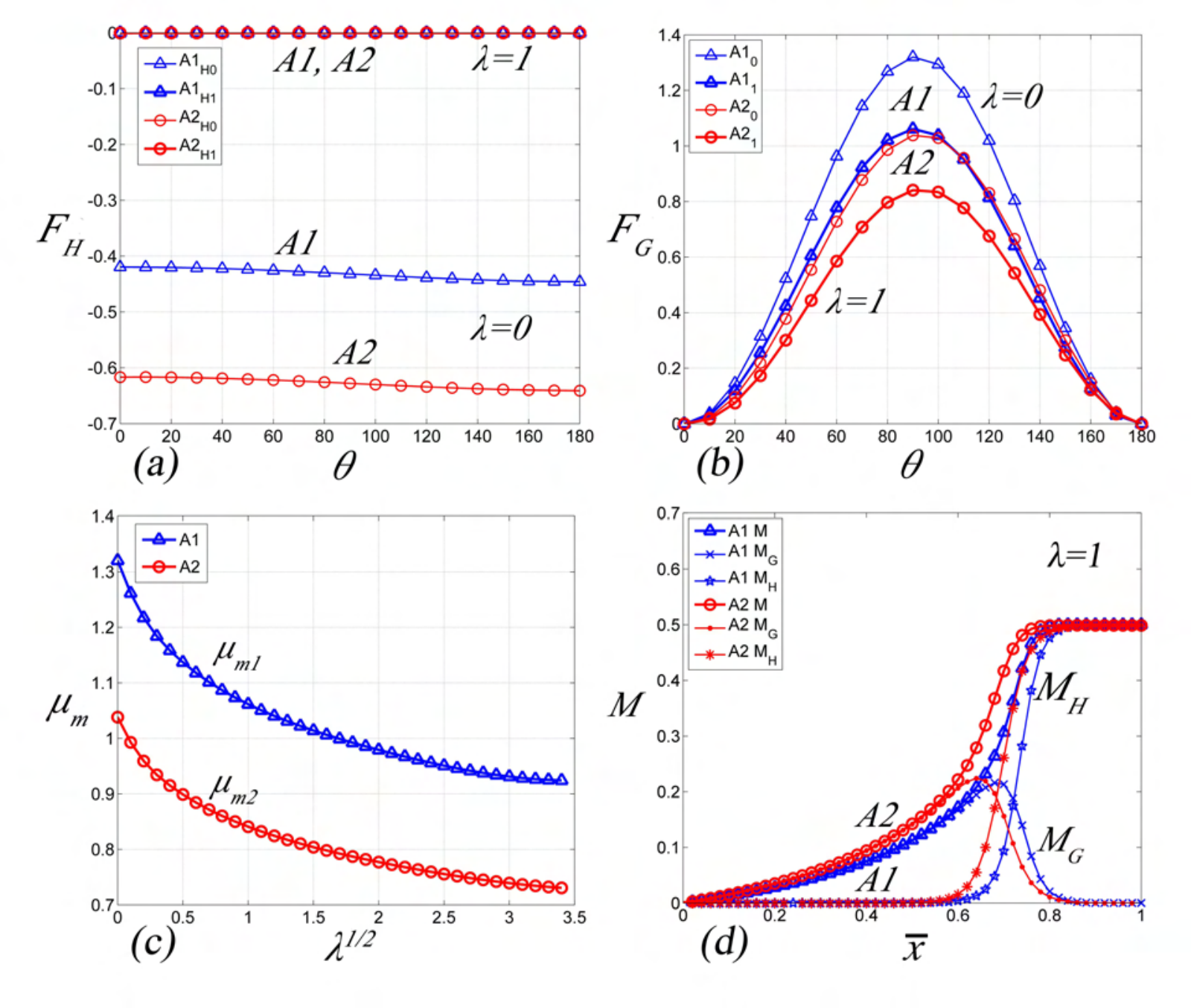} 
	\caption{Plots of (a) $F_{G}(\theta)$ and (b) $F_H(\theta)$  versus $\theta$ when, $\lambda=0$, 1. (c) Plots of $\mu_{m1}$ and $\mu_{m2}$ versus $\lambda^{1/2}$. (d) Plots of $M$, $M_G$ and $M_H$  versus $\bar{x}$ when $\lambda=1$.}
		\label{fig.4}
\end{figure}

From Maxwell electromagnetic theory, the 't Hooft's gauge potential of Eq. (\ref{eq.28}) at large $r$ tends to
\begin{eqnarray}
{\cal A}_i &=& (\cos\alpha + \cos\kappa)\partial_i\phi |_{r\rightarrow\infty} = \frac{\hat{\phi}_i}{r\sin\theta}\left(\frac{1}{2}(\cos\theta\pm1) + \frac{F_G(\theta)}{r}\right),
\label{eq.53}\\
F_G(\theta) &=& r\{h(P_1-\sin\theta) - g(P_2-\cos\theta) - \frac{1}{2}(\cos\theta \pm 1)\}|_{r\rightarrow\infty},
\label{eq.54}
\end{eqnarray}
for the Type $A$ and Type $B$ solutions respectively. From the graphs of $F_G(\theta)$ versus angle $\theta$, Figure \ref{fig.4} (a), we find that $F_G(\theta) =\mu_m\sin^2\theta$, where $\mu_m$ is the dimensionless magnetic dipole moment of the one-half monopole and is non vanishing for all values of $\lambda$. The value of $\mu_m$ is read from the graphs of $F_G(\theta)$ versus angle $\theta$ at $\theta=\frac{\pi}{2}$. The graphs of the magnetic dipole moment $\mu_{m1}$ and $\mu_{m2}$ of the Type $A$ solutions versus $\lambda^{1/2}$ when $g=\xi=1$ are shown in Figure \ref{fig.4} (c) for values of $0\leq\lambda\leq 12$ and the numerical values are given in Table \ref{table.1}. In the limit when $\lambda=0$, the dimensionless magnetic dipole moment for the Type 1 solution is $\mu_{m1}=\pm 1.32$ and for the Type 2 solution is $\mu_{m2}=\pm 1.04$. The plots of $\mu_{m1}$ and $\mu_{m2}$ versus $\lambda^{1/2}$ are non increasing graphs that concave upwards for both the Type $A$ solutions. Also noted is that, $\mu_{m1}>\mu_{m2}$, for the same value of $\lambda$. Similar findings are also recorded in Ref. \cite{kn:8} and \cite{kn:18} for the MAP solutions where the dimensionless magnetic dipole moment decreases from $\lambda=0$ ($\mu_m=4.72, 4.71$ respectively) to $\lambda=1$ ($\mu_m=3.25, 3.40$ respectively) when the electic charge parameter $\eta=0$.

At spatial infinity, these four one-half monopole solutions are exact and given by Eq. (\ref{eq.36}) to (\ref{eq.39}). From these exact solutions, we are able to conclude that the net magnetic charge of the four one-half monopole solutions is zero. This is because at spatial infinity, the Dirac string carries a magnetic flux of $\frac{2\pi}{g}$ going into the center of the sphere at infinity, $r=0$, and the one-half monopole at the origin carries a magnetic flux of $\frac{2\pi}{g}$ coming out from the origin, $r=0$. In order to check that our numerical calculations are consistent with the exact solutions, we calculate numerically and plot the graphs of the net magnetic charge $M$ (\ref{eq.8}), the topological magnetic charge $M_H$ (\ref{eq.9}), and the magnetic charge carried by the gauge field $M_G$ (\ref{eq.10}) versus the compactified axis, $\bar{x}=\frac{r}{r+1}$, for the Type 1 and 2 solutions when $\lambda=\xi=1$. See Figure \ref{fig.4} (d). We are only able to obtain the magnetic charge, $M_H$, of the one-half monopole from the numerical data. The numerical data cannot account for the magnetic charge carried by the Dirac string. We also notice from Figure \ref{fig.4} (d) that $M\rightarrow 0$ as $r\rightarrow 0$ and $M=\frac{1}{2}$ when $r\geq 4$. 

\begin{figure}[tbh]
	\centering
	\hskip0in
	 \includegraphics[width=5.2in,height=8.3in]{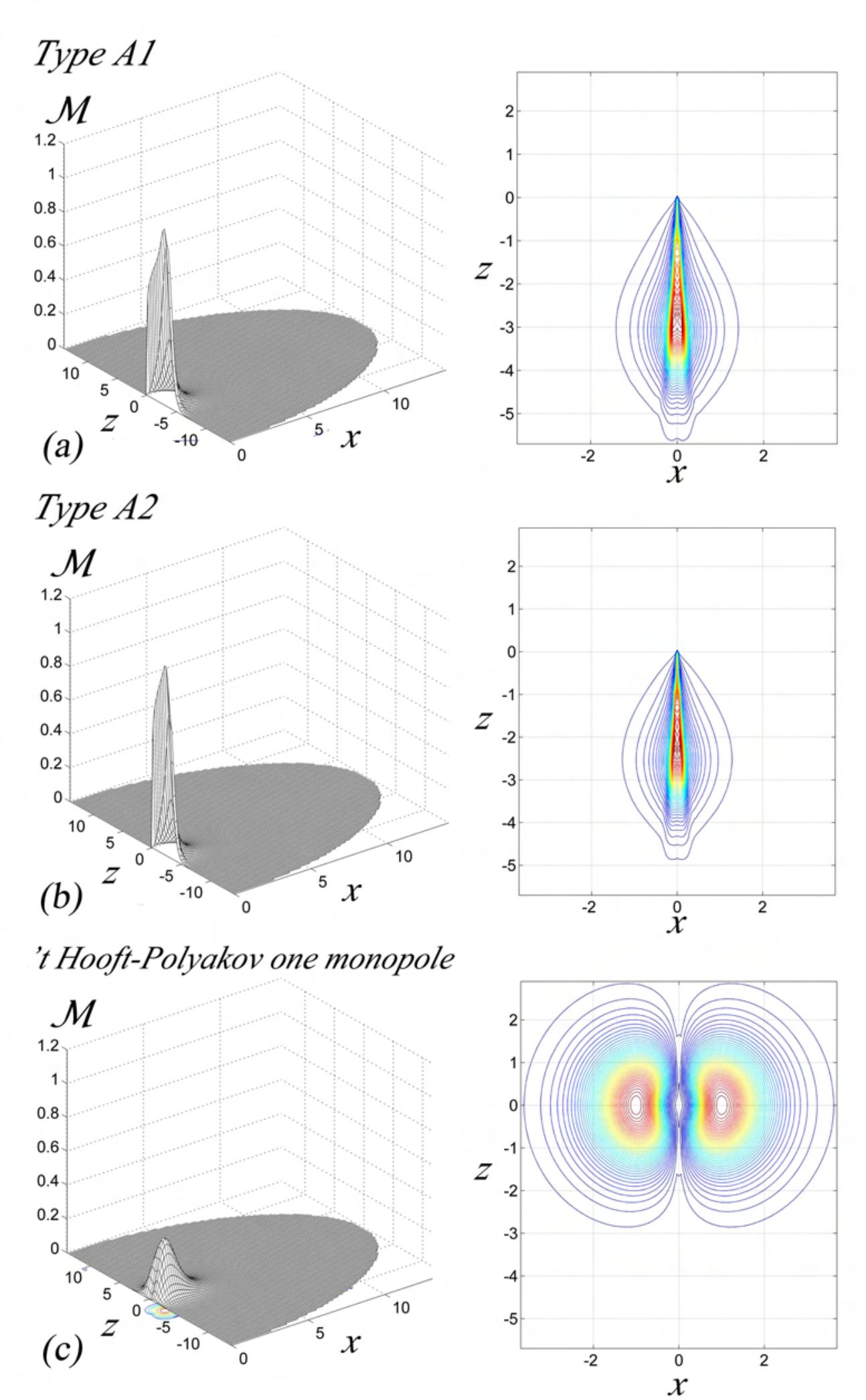} 
	\caption{3D surface and contour plots of the magnetic charge density, ${\cal M}$, of the (a) Type $A$1, (b) Type $A$2 and (c) the 't Hooft-Polyakov monopole solutions along the $x$-$z$ plane at $y=0$ when $\lambda=\xi=1$.}
	\label{fig.5}
\end{figure}

From Eq. (\ref{eq.8}), and using the definition for the Abelian magnetic field (\ref{eq.31}) given by Bogomol'nyi \cite{kn:3} and Faddeev \cite{kn:16}, we can write 
\begin{eqnarray}
M = \frac{1}{4\pi} \int \partial^i {\cal B}_i ~d^{3}x = \int{\cal M} ~d\theta~dr,  ~~\mbox{where} ~~{\cal M} = \frac{1}{2}r^2\sin\theta\{\partial^i {\cal B}_i\},
\label{eq.55}
\end{eqnarray}
is the magnetic charge density. The 3D surface and contour plots of ${\cal M}$ of (a) Type $A$1, (b) Type $A$2 solutions and (c) 't Hooft-Polyakov monopole the along the $x$-$z$ plane at $y=0$ when $\lambda=\xi=1$ are shown in Figure \ref{fig.5}. We note that the magnetic charge density is concentrated along the negative (positive) $z$-axis for the Type $A$ (Type $B$) solutions. We also note that the magnetic charge density is solely positive for the four one-half monopole solutions.


\begin{table}[tbh]
\begin{center}

\begin{tabular}{|c|c|c|c|c|c|c|c|c|c|c|c|}
\hline

$\lambda$ & 0 & 0.04 & 0.20 & 0.40 & 0.60 & 0.80 & 1.00 & 2.00 & 4.00 & 8.00 & 12.00 \\ \hline

$(\pm)\mu_{m1}$ & 1.32 & 1.22 & 1.15 & 1.11 & 1.09 & 1.07 & 1.06 & 1.02 & 0.98 & 0.94 & 0.92 \\ \hline

$(\pm)\mu_{m2}$ & 1.04 & 0.96 & 0.91 & 0.88 & 0.86 & 0.85 & 0.84 & 0.81 & 0.78 & 0.74 & 0.73 \\ \hline

$E_{1}$ & 0.51 & 0.54 & 0.56 & 0.57 & 0.58 & 0.59 & 0.59 & 0.61 & 0.63 & 0.66 & 0.67 \\ \hline

$E_{2}$ & 0.53 & 0.55 & 0.58 & 0.59 & 0.60 & 0.60 & 0.61 & 0.63 & 0.65 & 0.67 & 0.69 \\ \hline

\end{tabular}
\end{center}
\caption{The magnetic dipole moments $\mu_{m1}$, $\mu_{m2}$, and total energies $E_1$, $E_2$, of the Type 1 and 2 solutions respectively for $0\leq\lambda\leq 12$.}
\label{table.1}
\end{table}


\subsection{The Higgs Field and the Magnetic Field}
\label{subsection 5.4}

\begin{figure}[tbh]
	\centering
	\hskip-0.1in
	 \includegraphics[width=5.1in,height=8.2in]{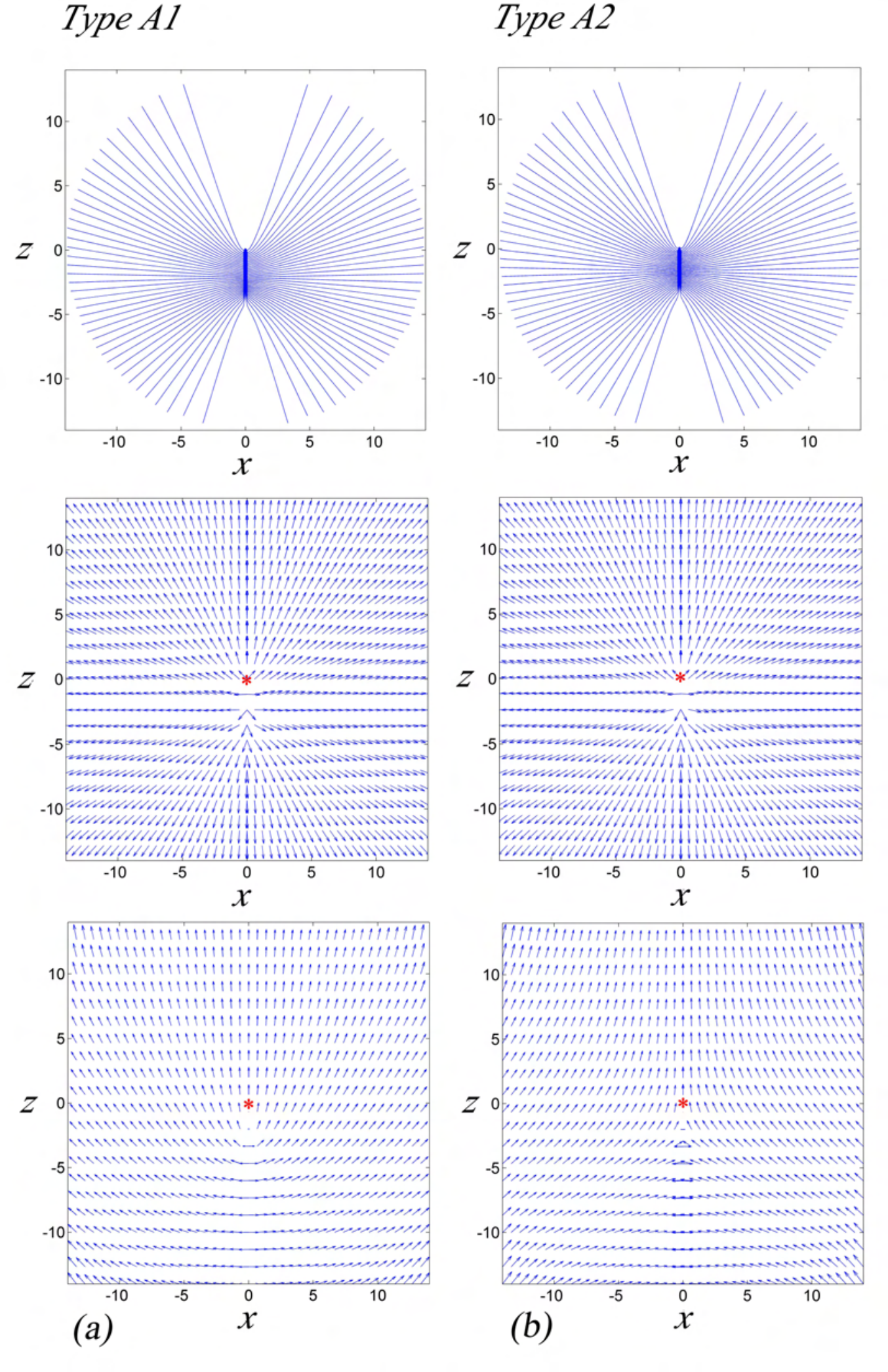} 
	\caption{Contour plots of the 't Hooft magnetic field lines, unit vector field plots of the 't Hooft magnetic field and Higgs field vector plots of the (a) Type $A$1 and (b) Type $A$2 solutions along the $x$-$z$ plane at $y=0$ when $\lambda=\xi=1$. The \textcolor{red}{*} indicates the location of the one-half monopole.}
	\label{fig.6}
\end{figure}

From Eq. (\ref{eq.28}), the 't Hooft's magnetic field lines contour plots of the Type $A$1 and Type $A$2 one-half monopole solutions along the $x$-$z$ plane at $y=0$ are shown in Figure \ref{fig.6} when $\lambda=\xi=1$. 
In the finite $r$ region near the origin, $r=0$, the magnetic field lines of the numerical Type $A$ (Type $B$) one-half monopole are string-like and concentrated along a finite length of the negative (positive) $z$-axis before they become hedgehog-like at large $r$. For the Type 1 solutions, the magnetic field line string stretches from $z\approx \pm 0.14$ to $z\approx\mp 3.82$ and for the Type 2 solutions, the magnetic field line string stretches from $z\approx \pm 0.14$ to $z\approx\mp 3.25$.

The direction of the magnetic field lines of the one-half monopole solutions are shown by the vector field plots of the 't Hooft's magnetic field unit vector $\hat{B}_i$ (\ref{eq.29}) along the $x$-$z$ plane at $y=0$ when $\lambda=\xi=1$. See Figure \ref{fig.6}. The asterisk \textcolor{red}{*} indicates the location of the one-half monopole at $r=0$. The presence of a one-half monopole sitting at the origin is once again indicated in Figure \ref{fig.6} by the location of the Higgs field's node marked asterisk \textcolor{red}{*} on the Higgs field's vector, $\Phi^a$, plots along the $x$-$z$ plane at $y=0$. 

\begin{figure}[tbh]
	\centering
	\hskip-0.1in
	 \includegraphics[width=5.1in,height=8.3in]{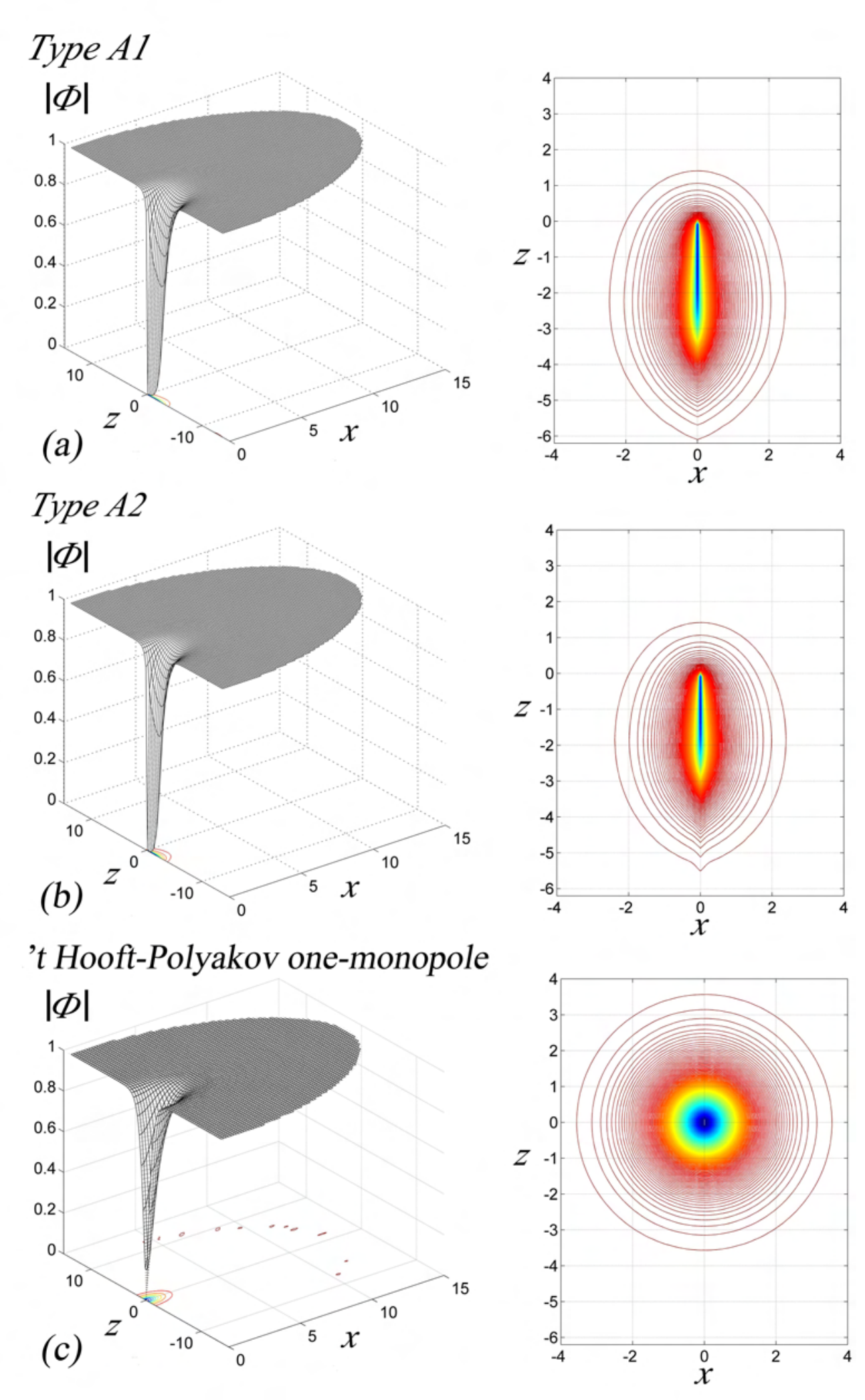} 
	\caption{The 3D surface and contour plots of the Higgs field modulus $|\Phi|$ of the (a) Type $A$1, (b) Type $A$2 and (c) the 't Hooft-Polyakov monopole solutions along the $x$-$z$ plane at $y=0$ when $\lambda=\xi=1$.}
	\label{fig.7}
\end{figure}

The 3D surface and contour plots of the modulus of the Higgs field functions $|\Phi|$ of the (a) Type $A$1 and (b) Type $A$2 one-half monopole solutions and (c) the 't Hooft-Polyakov monopole solution along the $x$-$z$ plane at $y=0$ when $\lambda=\xi=1$, show that there is a point zero of the Higgs field modulus at $r=0$ as shown in Figure \ref{fig.7}. Hence the one-half monopoles and the one-monopole are located at $r=0$ where the Higgs field vanishes. The shape of the 3D surface plots for the one-half monopole in Figure \ref{fig.7} (a) and (b) is that of a flatten cone with its half-rounded vertex touching $r=0$. The cones are flatten along the negative $z$-axis for the Type $A$ one-half monopole. This is in contrast to the 3D surface plot of the Higgs modulus for the 't Hooft-Polyakov monopole which is a sharp vertex circular cone with its vertex at $r=0$, where there is only one zero of $|\Phi|$ at $r=0$. In the MAP solutions, there is also only one zero of $|\Phi|$ at the location of the MAP dipole when the $\phi$-winding number, $n=1$. However when $n=2$, there is a double zero of $|\Phi|$ at the location of the MAP dipole along the $z$-axis in the direction perpendicular to the $z$-axis \cite{kn:18}. In the case of the one-half monopole there is a double zero of $|\Phi|$ at $r=0$ along one side of the $z$-axis only, the negative (positive) $z$-axis for the Type $A$ (Type $B$) solutions.
 
The modulus of the Higgs field at large $r$ tends to
\begin{eqnarray}
g|\Phi(r,\theta)|_{r\rightarrow\infty} = (\xi - \frac{F_H(\theta)}{r}). 
\label{eq.56}
\end{eqnarray}
From the graphs of the Higgs function $F_H(\theta)$ versus the angle $\theta$, Figure \ref{fig.4} (b), we find that $F_H(\theta)$ is a non vanishing constant $c_1$ only when $\lambda=0$ for both the Type 1 and Type 2 solutions. The constant $c_1=-0.43$ for the Type 1 solutions and $c_1=-0.63$ for the Type 2 solutions. 
For non vanishing values of $\lambda$, the constant $c_1$ vanishes for both Type 1 and Type 2 solutions.

\subsection{The Energy Density and Total Energy}
\label{subsection 5.5}

\begin{figure}[tbh]
	\centering
	\hskip0in
	 \includegraphics[width=5.8in,height=8.3in]{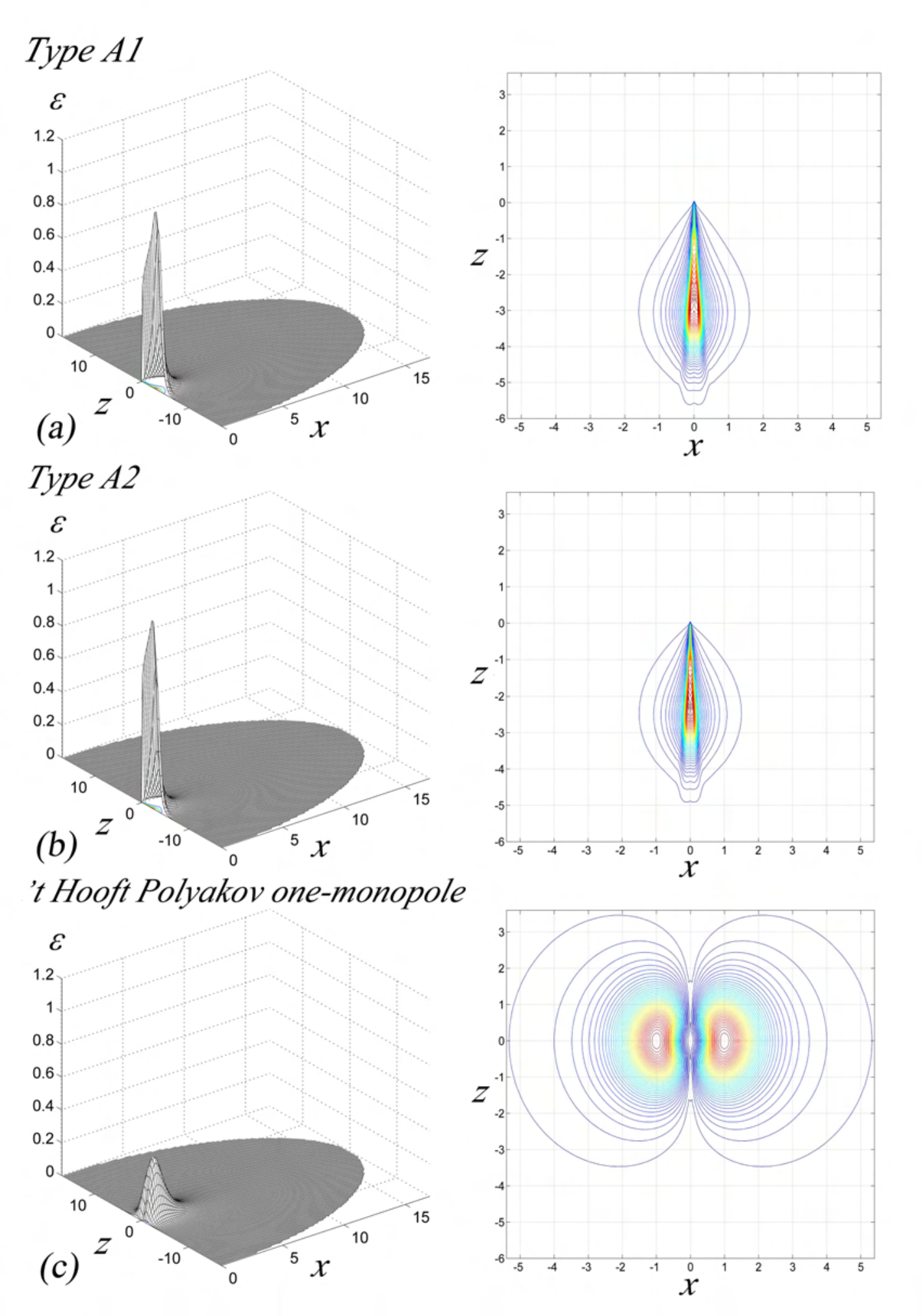} 
	\caption{3D surface and contour plots of the energy density ${\cal E}$ of the (a) Type $A$1, (b) Type $A$2 and (c) the 't Hooft-Polyakov monopole solutions along the $x$-$z$ plane at $y=0$ when $\lambda=\xi=1$.}
	\label{fig.8}
\end{figure}

The total dimensionless energy (\ref{eq.12}) when $\xi$=$g$=1, can be written as
\begin{eqnarray}
E = \int{\cal E} d\theta~dr, ~~{\cal E} = \frac{1}{4}r^2\sin\theta\{B^a_iB^a_i +  D_i\Phi^aD_i\Phi^a + \frac{\lambda}{2}(\Phi^a\Phi^a-\xi^2)^2\},
\label{eq.57}
\end{eqnarray}
where ${\cal E}$ is the energy density.
The 3D surface and contour line plots of the energy density ${\cal E}$
of the (a) Type $A$1, (b) Type $A$2 one-half monopole solutions and (c) the 't Hooft-Polyakov monopole solution along the $x$-$z$ plane at $y=0$ when $\lambda=1$ are shown in Figure \ref{fig.8}. The Type 1 solutions has a slightly wider spread of the energy distribution along one-half of the $z$-axis and a slightly lower energy density peak with the same value of $\lambda=1$ compared to the Type 2 solutions. For the Type $A$1 solution the energy density peak is located along the $z$-axis at $-2.941$ with peak value of 1.087 while the energy density peak of the Type $A$2 solution is located along the $z$-axis at $z=-2.235$ with peak value of 1.149. The energy density concentration of the Type $A$ (Type $B$) solutions are along the negative (positive) $z$-axis centered around its peak value. The 't Hooft-Polyakov monopole solution has two peak values of 0.3629 at $z$=0 and $x=\pm 0.9412$. Hence the shape of the one-half monopole is that of a rugby ball whereas the 't Hooft-Polyakov monopole is shaped like a torus.

\begin{figure}[tbh]
	\centering
	\hskip-0.5in
	 \includegraphics[width=5.8in,height=5.5in]{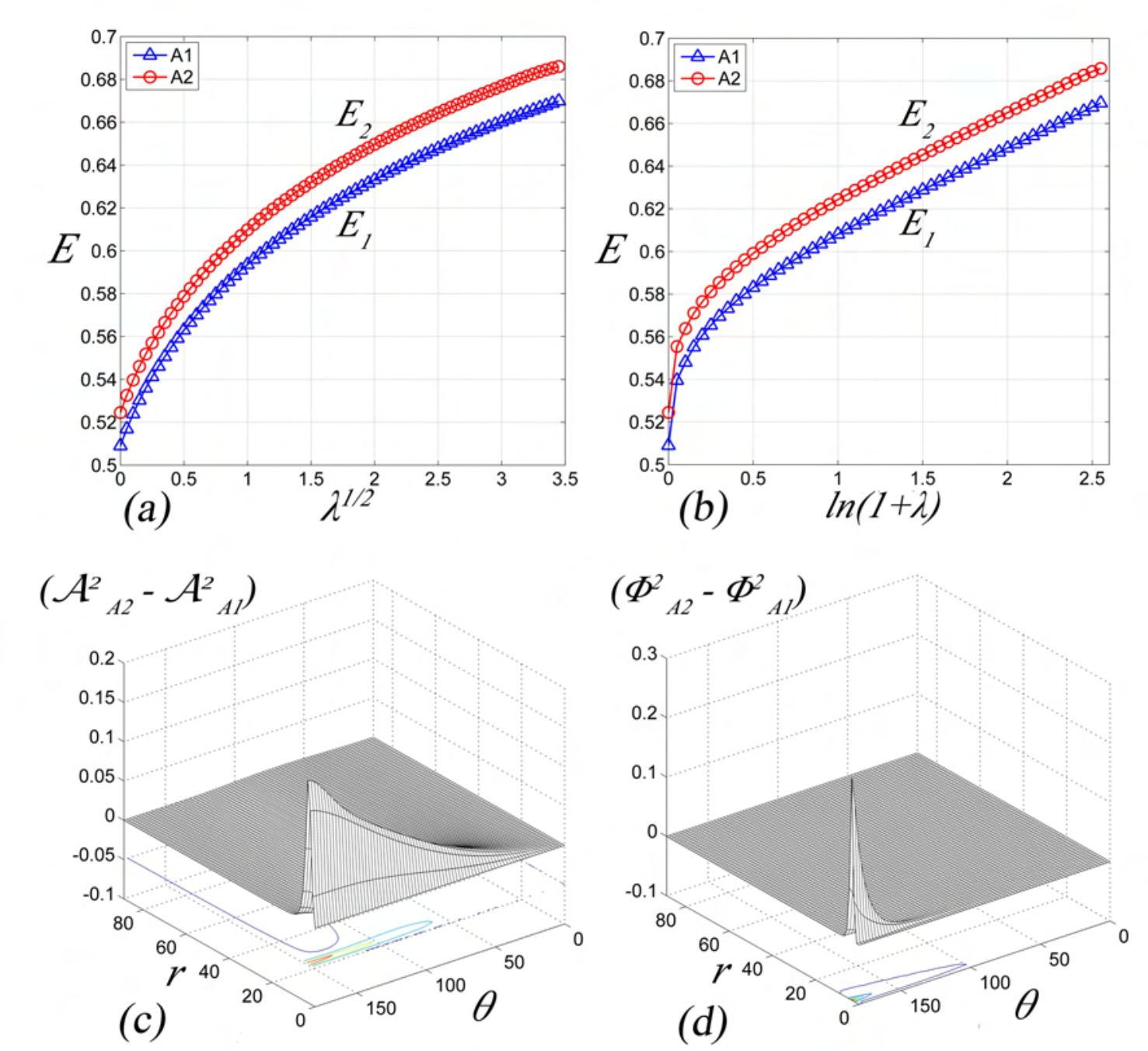} 
	\caption{Plots of $E_1$ and $E_2$ versus (a) $\lambda^{1/2}$ and (b) $\ln(1+\lambda)$. Here $g=\xi=1$. The 3D surface plots of (c) (${\cal A}^2_{A2} - {\cal A}^2_{A1}$) and (d) ($\Phi^2_{A2} - \Phi^2_{A1}$) when $\lambda=1$.}
	\label{fig.9}
\end{figure}

The total dimensionless energy (\ref{eq.57}) of the Type $1$ one-half monopole solutions is $E_1=0.509$ and that of the Type $2$ solutions is $E_2=0.525$ when $\lambda=0$. 
Their total energies seem to be higher than the BPS total energy of a one-half monopole which is $\frac{1}{2}$. The total energies of the one-half monopole solutions are plotted versus $\lambda^{1/2}$ and $\ln(1+\lambda)$ for values of Higgs field strength $0\leq \lambda\leq 12$, Figure \ref{fig.9} (a) and (b). For a particular value of $\lambda$, the Type 2 solutions possess higher total energy, compared to the Type 1 solutions, $E_1>E_2$. The total  energy $E_1$ and $E_2$ versus $\lambda^{1/2}$ curves are non decreasing graphs that concave downwards, Figure \ref{fig.9} (a), whereas the total energy $E_1$ and $E_2$ versus $\ln(1+\lambda)$ graphs become linear when $\lambda$ is large, Figure \ref{fig.9} (b).
The table of total energies $E_1$ and $E_2$ for different values of $\lambda$ of the one-half monopole solutions are shown in Table \ref{table.1}. 

\subsection{Comparison between the Numerical Solutions}
\label{subsection 5.6}

The boundary condition at $r\rightarrow \infty$ is fixed by the exact one-half monopole solutions (\ref{eq.36}) - (\ref{eq.39}) and the four exact solutions are related by gauge transformations. However near the origin as $r\rightarrow 0$ and along the positive and negative $z$-axis (or at $\theta=0$ and $\pi$ respectively), the solutions are fixed only by similar boundary conditions. A direct method of checking the similarities between the Type $A$ and Type $B$ solutions is to rotate the $z$-axis of the Type $A$ solution about its center at $r=0$, $180^o$ so the the positive $z$-axis becomes the negative $z$-axis. The transformation performed on the Type $A$ profile functions, $P_1$, $\psi_1$, and $\Phi_1$, to make it similar to the respective Type $B$ profile functions is 
\begin{equation}
F_A(r,\theta) \rightarrow F_A(r, \pi - \theta) = \tilde{F}_A(r,\theta) \approx F_B(r,\theta),
\label{eq.58}
\end{equation}
and the transformation performed on the Type $A$ profile functions, $P_2$, $R_1$, and $\Phi_2$, to make it similar to the respective Type $B$ profile functions is 
\begin{equation}
F_A(r,\theta) \rightarrow - F_A(r, \pi - \theta) = \tilde{F}_A(r,\theta) \approx F_B(r,\theta),
\label{eq.59}
\end{equation}
where $F_A(r,\theta)$ and $F_B(r,\theta)$ represent the Type $A$ and Type $B$ profile functions respectively. The transformed Type $A$1(2) profile functions, $\tilde{F}_A(r,\theta)$ are then subtracted from the respective Type $B$1(2) profile functions, $F_B(r,\theta)$. In our numerical calculations, we found that the 3D surface plots of the differences, $F_B(r,\theta)-\tilde{F}_A(r,\theta)$, versus $r$ and $\theta$ is very small. They are of the order of $10^{-13}$ for the Type 1 solutions and of the order of $10^{-12}$ for the Type 2 solutions. Hence the Type $B$ solutions are exactly $180^o$ rotations of the $z$-axis about $r=0$ of the respective Type $A$ solutions.

In order to compare the four types of solutions with each other to check if the Type 1 and Type 2 solutions are genuinely different and also to confirm that the Type $A$ and Type $B$ solutions are gauge equivalent, we used the gauge invariant form of the gauge potential and the Higgs field,
\begin{equation}
{\cal A}^2=r^2\sin^2\theta A^a_i A^a_i, ~~\Phi^2 = \Phi^a\Phi^a.
\label{eq.60}
\end{equation}
From the four types of solutions, there are six combinations of the differences between the solutions of the quantities ${\cal A}^2$ and $\Phi^2$. We then plot twelve 3D surface graphs of the differences of ${\cal A}^2$ and $\Phi^2$ versus $r$ and $\theta$ for all the six combinations. We find that for the graphs of ($\tilde{{\cal A}}^2_{A1} - {\cal A}^2_{B1}$), ($\tilde{\Phi}^2_{A1} - \Phi^2_{B1}$), ($\tilde{{\cal A}}^2_{A2} - {\cal A}^2_{B2}$), and ($\tilde{\Phi}^2_{A2} - \Phi^2_{B2}$), versus $r$ and $\theta$, the order of magnitudes are very small, $10^{-12}$, $10^{-13}$, $10^{-15}$, and $10^{-15}$ respectively. Therefore the Type $A$ and Type $B$ solutions are gauge equivalent solutions and they differ by only a rotation of $180^o$ of the $z$-axis about its center in 3D space. The transformed functions $\tilde{{\cal A}}^2$ and $\tilde{\Phi}^2$ obey the transformation (\ref{eq.58}).

However for the 3D graphs of ($\tilde{{\cal A}}^2_{A2} - {\cal A}^2_{B1}$), ($\tilde{\Phi}^2_{A2} - \Phi^2_{B1}$), ($\tilde{{\cal A}}^2_{A1} - {\cal A}^2_{B2}$), ($\tilde{\Phi}^2_{A1} - \Phi^2_{B2}$), (${\cal A}^2_{A2} - {\cal A}^2_{A1}$), ($\Phi^2_{A2} - \Phi^2_{A1}$), (${\cal A}^2_{B2} - {\cal A}^2_{B1}$), and ($\Phi^2_{B2} - \Phi^2_{B1}$) versus $r$ and $\theta$, there are significant differences between the Type 1 and Type 2 solutions along the $z$-axis in the region where the energy of the one-half monopole is concentrated. As expected the four sets of graphs for the differences in ${\cal A}^2$ and  ${\Phi}^2$ are similar. We therefore show in Figure \ref{fig.9} (c) and (d) only one of the four sets, the 3D surface plot of (${\cal A}^2_{A2} - {\cal A}^2_{A1}$) and ($\Phi^2_{A2} - \Phi^2_{A1}$) when $\lambda=1$. The maximum value of (${\cal A}^2_{A2} - {\cal A}^2_{A1}$) is 0.1825  and  ($\Phi^2_{A2} - \Phi^2_{A1}$) is 0.2750. Both maxima occur at the same point along one side of the $z$-axis at $r=3.38$ and $\theta=\pi$. Hence the Type 1 and Type 2 solutions are not gauge equivalent one-half monopole. They are different and the differences in the solutions occur in the region of where the energy of the solutions is concentrated.

We also note that the gauge transformations (\ref{eq.19}) connecting the Type $A$1 (\ref{eq.36}) to $B$1 (\ref{eq.38}) solutions and the gauge transformation connecting the Type $A$2 (\ref{eq.37}) to $B$2 (\ref{eq.39}) solutions  are. 
\begin{equation}
f_{(A1\rightarrow B1)}=-\frac{\pi}{2}, ~~f_{(A2\rightarrow B2)}=\frac{\pi}{2}.
\label{eq.61}
\end{equation}
The gauge transformations (\ref{eq.19}) connecting the Type $A$1 (\ref{eq.36}) to $B$2 (\ref{eq.39}) solutions and that connecting the Type $B$1 (\ref{eq.38}) to $A$2 (\ref{eq.37}) solutions are
\begin{equation}
f_{(A1\rightarrow B2)}=f_{(B1\rightarrow A2)}=\theta+\frac{\pi}{2}.
\label{eq.62}
\end{equation}
The gauge transformations (\ref{eq.19}) connecting the Type $A$1 (\ref{eq.36}) to $A$2 (\ref{eq.37}) solutions and that connecting the Type $B$1 (\ref{eq.38}) to $B$2 (\ref{eq.39}) solutions are
\begin{equation}
f_{(A1\rightarrow A2)}= \theta, ~~~f_{(B1\rightarrow B2)}=\theta+\pi.
\label{eq.63}
\end{equation}
The gauge transformations (\ref{eq.61}) are trivial as $f$ is a constant whereas the gauge transformations (\ref{eq.62}) and (\ref{eq.63}) are non-trivial and depend on $\theta$.


\section{Comments}

We have found four types of one-half monopole solutions, Type $A$1, Type $A$2, Type $B$1, and Type $B$2 solutions. All these four solutions possess zero net magnetic charge. By comparing all the six profile functions of the Type $A$ and Type $B$ solutions, we are able to conclude that the Type $B$ solutions are exact $180^o$ rotation of the $z$-axis about $r=0$ of the respective Type $A$ solutions in 3D space. Again by comparing the gauge invariant form of the gauge potentials, ${\cal A}^2$, and the Higgs fields, $\Phi^2$, we are able to conclude that the Type $A$ and Type $B$ solutions are gauge equivalent. However the Type 1 and Type 2 solutions are significantly distinct in the region where the energy of the one-half monopole is concentrated. The difference in the Type 1 and Type 2 solutions are also reflected in the graphs of magnetic dipole moments versus $\lambda^{1/2}$, Figure \ref{fig.4} (c); and the graphs of total energy versus $\lambda^{1/2}$ and $\ln(1+\lambda)$, Figure \ref{fig.9} (a) and (b).
All the one-half monopole solutions reported here satisfy the gauge condition, $r{\partial_r}R_1 - {\partial_\theta}\psi_1 = 0$ \cite{kn:8}. However this gauge condition still do not ensure the uniqueness of solutions for these one-half monopole configurations. We have in fact found two distinct one-half monopole, the Type 1 and Type 2 one-half monopole, which are connected by the non trivial gauge transformations (\ref{eq.62}) and (\ref{eq.63}).

For further work on this subject, we would like to introduce electric charge into the Type 1 and Type 2 one-half monopole solutions using the standard procedure of Julia and Zee \cite{kn:19} by letting the time component of the gauge potential at large distances be nonzero and introducing a constant electric charge parameter into the exact asymptotic solutions at large $r$. The numerical finite energy rotating one-half dyon solutions will be presented in a separate work soon. 

Another direction of our further work is to look for a one-monopole and a one-half monopole of opposite magnetic charge in the SU(2) YMH theory. We have already found these type of solutions and will be finalizing our findings in a separate work soon.


\section{Acknowlegements}
The authors would like to thank Universiti Sains Malaysia for the RU research grant (account number: 1001/PFIZIK/811180).


\begin{thebibliography}{99}

\bibitem[1]{kn:1} G. 't Hooft, Nucl. Phy. {\bf B79},  276 (1974); 
A.M. Polyakov, Sov. Phys. - JETP {\bf 41}, 988 (1975); Phys. Lett. {\bf B59},  82 (1975); JETP Lett. {\bf 20}, 194 (1974).

\bibitem[2]{kn:2} E.B. Bogomol'nyi and M.S. Marinov, Sov. J. Nucl. Phys. {\bf 23}, 357 (1976).

\bibitem[3]{kn:3} M.K. Prasad and C.M. Sommerfield, Phys. Rev. Lett. {\bf 35}, 760 (1975);
E.B. Bogomol'nyi, Sov. J. Nucl. Phys. {\bf 24}, 449 (1976).

\bibitem[4]{kn:4} C. Rebbi and P. Rossi, Phys. Rev. {\bf D22}, 2010 (1980);
R.S. Ward, Commun. Math. Phys. {\bf 79}, 317 (1981);
P. Forgacs, Z. Horvarth and L. Palla, Phys. Lett. {\bf B99}, 232 (1981); 
Nucl. Phys. {\bf B192}, 141 (1981);
M.K. Prasad, Commun. Math. Phys. {\bf 80}, 137 (1981); 
M.K. Prasad and P. Rossi, Phys. Rev. {\bf D24}, 2182 (1981).

\bibitem[5]{kn:5} E.J. Weinberg and A.H. Guth, Phys. Rev. {\bf D14}, 1660 (1976).

\bibitem[6]{kn:6} Rosy Teh and K.M. Wong, J. Math. Phys. {\bf 46}, 082301 (2005); Int. J. Mod. Phys. {\bf A20}, 4291 (2005).

\bibitem[7]{kn:7} P.M. Sutcliffe, Int. J. Mod. Phys. {\bf A12}, 4663 (1997);
C.J. Houghton, N.S. Manton and P.M. Sutcliffe, Nucl.Phys. {\bf B510}, 507 (1998).

\bibitem[8]{kn:8} B. Kleihaus and J. Kunz, Phys. Rev. {\bf D61}, 025003 (2000);
B. Kleihaus, J. Kunz, and Y. Shnir, Phys. Lett. {\bf B570}, 237, (2003); 
B. Kleihaus, J. Kunz, and Y. Shnir, Phys. Rev. {\bf D68}, 101701 (2003); Phys. Rev. {\bf D70}, 065010 (2004).

\bibitem[9]{kn:9} Rosy Teh, K.G. Lim and P.W. Koh, {\it Magnetic Half-Monopole Solutions}, FRONTIERS IN PHYSICS: 3rd International Meeting, Kuala Lumpur (Malaysia), 12-16 January 2009, edited by S.P. Chia, M.R. Muhammad, and K. Ratnavelu,  ISBN: 978-0-7354-0687-2, AIP Conference Proceedings Volume {\bf 1150}, 424 (2009).

\bibitem[10]{kn:10} E. Harikumar, I. Mitra, and H.S. Sharatchandra, Phys. Lett. {\bf B557}, 303 (2003).

\bibitem[11]{kn:11} Rosy Teh and K.M. Wong, {\it Half-Monopole and Multimonopole}, Int. J. Mod. Phys. {\bf A20}, 2195 (2005).

\bibitem[12]{kn:12} N.S. Manton, Nucl. Phys. (N.Y.) {\bf B126}, 525 (1977).

\bibitem[13]{kn:13} J. Arafune, P.G.O. Freund, and C.J. Goebel, J. Math. Phys. {\bf 16}, 433 (1975).

\bibitem[14]{kn:14} A. Actor, Rev. Mod. Phys. {\bf 51}, 461 (1979).

\bibitem[15]{kn:15} S. Coleman, {\it New Phenomena in Subnuclear Physics}, Proc. 1975 Int. School of Physics `Ettore Majorana', ed A Zichichi, New York Plenum, 297 (1975).

\bibitem[16]{kn:16} L.D. Faddeev, {\it Nonlocal, Nonlinear and Nonrenormalisable Field Theories}, Proc. Int. Symp., Alushta, Dubna: Joint Institute for Nuclear Research, 207 (1976); Lett. Math. Phys. {\bf 1}, 289 (1976).


\bibitem[17]{kn:17} D.G. Boulware et al., Phys. Rev. {\bf D14}, 2708 (1976). 

\bibitem[18]{kn:18} K.G. Lim, Rosy Teh and K.M. Wong, J. Phys. G: Nucl. Part. Phys. {\bf 39}, 025002 (2012).


\bibitem[19]{kn:19} B. Julia and A. Zee, Phys. Rev. {\bf D11}, 2227 (1975).

\end{thebibliography}
\end{document}